\documentclass[smallextended]{svjour3}

\usepackage{listings}
\usepackage{xcolor}
\usepackage{multirow}
\usepackage{fancybox}
\usepackage{framed}
\usepackage{tabularx}
\usepackage{makecell}
\usepackage{xspace}
\usepackage[table]{xcolor}
\usepackage{booktabs}
\usepackage{array}
\usepackage{graphicx}
\usepackage{pifont}
\usepackage{url}
\usepackage{xurl}
\usepackage{hyperref}

\newcommand{\xmark}{\ding{55}}
\newcommand{\pmark}{\raise.17ex\hbox{$\scriptstyle\mathtt{\sim}$}}

\definecolor{darkgreen}{RGB}{0,100,0}
\definecolor{darkpurple}{RGB}{64,0,96}

\lstdefinelanguage{json}{
    morestring=[b]",
    morecomment=[l]{//},
    morekeywords={true, false, null},
    sensitive=false
}
\makeatletter
\def\subsubsection{\@startsection{subsubsection}{3}{\z@}%
  {-3.25ex\@plus -1ex \@minus -.2ex}%
  {1ex \@plus .2ex}%
  {\normalfont\normalsize\bfseries}}
\makeatother

\journalname{Noname}

\newcommand{\rqone}{How compliant and complete are GitHub SBOMs?}
\newcommand{\rqtwo}{How consistent is the GitHub SBOM Tool compared to other SBOM generators?}
\newcommand{\rqthree}{How useful is the GitHub SBOM Tool compared to other SBOM generators in vulnerability tracking?}

\newcommand{\ghtool}{GitHub SBOM Tool\xspace}
\newcommand{\mstool}{Microsoft SBOM Tool\xspace}
\newcommand{\trivy}{Trivy\xspace}
\newcommand{\syft}{Syft\xspace}

\begin{document}

\title{Beyond Compliance: A Large Scale Study on the Completeness and Consistency of the GitHub SBOMs}

\titlerunning{Beyond Compliance: A Large Scale Study on the Completeness and Consistency of the GitHub SBOMs}

\author{
  Kawsar Ahmed Bhuiyan \and
  Mohamed Bilel Besbes \and
  Rachna Raj \and
  Adam Al Assil \and
  Diego Elias Costa
}

\authorrunning{Bhuiyan et al.}

\institute{
  Kawsar Ahmed Bhuiyan \at
    REALISE Lab, Concordia University, Montréal, Canada \\
    \email{kawsarahmed.bhuiyan@mail.concordia.ca}
  \and
  Mohamed Bilel Besbes \at
    REALISE Lab, Concordia University, Montréal, Canada \\
    \email{mohamedbilel.besbes@mail.concordia.ca}
  \and
  Rachna Raj \at
    REALISE Lab, Concordia University, Montréal, Canada \\
    \email{rachna.raj@mail.concordia.ca}
  \and
  Adam Al Assil \at
    REALISE Lab, Concordia University, Montréal, Canada \\
    \email{adam.alassil@concordia.ca}
  \and
  Diego Elias Costa \at
    REALISE Lab, Concordia University, Montréal, Canada \\
    \email{diego.costa@concordia.ca}
}

\date{Received: date / Accepted: date}

\maketitle

\begin{abstract}
Modern software development relies heavily on open-source components. Reusing components accelerates innovation but increases exposure to supply-chain attacks exploiting known vulnerabilities. Software Bills of Materials (SBOMs) improve software supply chain transparency by enumerating components, their versions, and their provenance. GitHub, the largest open-source development hosting platform, now automatically generates SBOMs for repositories, providing valuable metadata for risk assessment. Yet, it is unclear whether GitHub SBOMs can serve as a reliable source for vulnerability and license analysis, and how incomplete or inconsistent metadata may affect different programming ecosystems. To address this, we conduct a large-scale analysis of 10,000 GitHub repositories across ten programming language ecosystems, evaluating GitHub SBOMs against three other popular SBOM generators: Syft, Trivy, and the Microsoft SBOM Tool. Our study finds a lack of NTIA compliance in GitHub SBOMs, though core metadata is consistently present. We also find that component version and license information availability is highly dependent on the programming ecosystem. Compared with the other three tools, GitHub yields results similar to the Microsoft SBOM Tool and often outperforms Syft and Trivy in providing version and license information. Finally, we discuss potential shortcomings of the GitHub SBOM Tool, directly related to how each ecosystem manages its dependencies.

\end{abstract}

\section{Introduction}

The increasing reliance on third-party and open-source software components has become a defining characteristic of contemporary software development~\cite{wang2020empirical}. 
While this trend enables faster innovation and reduces engineering costs, it also increases the risk of exploitation, as attackers may target known vulnerabilities in these components, potentially undermining system integrity, exposing sensitive data, or causing significant operational and economic damage \cite{Ladisa2023}\cite{Enck2022}\cite{Ohm2020}. 
These challenges are intensified by the growing scale and complexity of software dependency networks, which often span thousands of interrelated packages \cite{technicalLeverage2021}\cite{Samaana_2025}.

To address these concerns, the Software Bill of Materials (SBOM) has emerged as a key mechanism for enhancing transparency in the software supply chain. An SBOM provides a structured inventory of all software components, including open-source and proprietary modules, along with their dependencies and relationships~\cite{linuxfoundation_sbom_2021}~\cite{Xia2024Trust}. By making this information transparent, SBOMs allow downstream users to assess potential cybersecurity and licensing risks~\cite{cisa_sbom_faq_2024}~\cite{github_sbom_article_2025}. The strategic importance of SBOMs was recognized in the U.S. Executive Order on Improving the Nation’s Cybersecurity, issued in May 2021, which mandated the use of SBOMs in federal software procurement~\cite{us2021eo14028}. To promote standardization, the National Telecommunications and Information Administration (NTIA) established the Software Transparency initiative, defining guidelines for machine-readable SBOMs that specify minimum required elements such as supplier and component names, version information, dependency relationships, and other essential metadata~\cite{ntia2021minimum}. These guidelines aim to harmonize SBOM practices across organizations, enabling scalable and interoperable adoption.

Within this context, GitHub has emerged as a significant platform for SBOM generation and dissemination. GitHub hosts millions of open-source repositories and provides native support for SBOM creation by leveraging its dependency graph \cite{githubSbomExport2025}. 
Hence, GitHub is a compelling target for empirical investigation, particularly given its ecosystem diversity and scale.
As users may adopt GitHub SBOMs as part of their compliance and quality assessment~\cite{Nocera2023Software}\cite{sbomsgainingtraction}, it is crucial to assess their compliance and consistency and to compare them with other widely used SBOM generation tools. 

In this paper, we present a large-scale empirical study of GitHub-generated SBOMs across 10 popular programming language ecosystems: C, C++, C\#, Python, PHP, Java, JavaScript, Go, Swift, and Rust.
We collect SBOM data from 10,000 GitHub repositories, and design our study to answer the following research questions: 

\textbf{RQ1: \rqone}
We examine whether GitHub SBOMs include the key metadata required for software supply chain activities, focusing on the compliance to NTIA minimum elements.
We then evaluate the presence of version and license information of GitHub SBOMs' components. 
We found that GitHub SBOMs include most core metadata, such as component names, unique identifiers, and dependency relationships, but none fully meet NTIA minimum requirements due to missing supplier information. Version information is generally available for top-level dependencies, while transitive dependencies are only listed when their parent dependencies include versions. License information is inconsistently reported across dependency components.

\textbf{RQ2: \rqtwo}
This question explores how GitHub SBOMs differ from those produced by three widely used tools in terms of metadata coverage and consistency across ecosystems.
We found that GitHub SBOMs generally report more dependencies and provide version and license information more frequently than those generated by Trivy and Syft. 
However, SBOMs generated by the \mstool tend to more consistently include version information. 

\textbf{RQ3: \rqthree}
This question evaluates how the \ghtool compares against other SBOM tools in supporting automated vulnerability tracking. 
We found that \ghtool and \mstool tend to list more components, and more frequently include unique identifiers (PURL) that can be used for vulnerability tracking.
\syft and \trivy, on the other hand, frequently include components that lack precise version information.
Consequently, GitHub SBOMs can be mapped to more vulnerabilities than other tools' SBOMs across nearly all languages. However, the accuracy of vulnerability detection depends on whether exact version information is provided for the components.

The main contributions of our paper are as follows:
\vspace{-5pt}
\begin{itemize}   
    \item We conduct the first comprehensive empirical study of GitHub-generated SBOMs using 10,000 repositories (1,000 repositories per programming language) across 10 programming languages, evaluating SBOM quality and completeness against the NTIA minimum element requirements.
    
    \item We compare four major SBOM generators, revealing variations in version coverage, availability of unique identifiers, and license reporting across programming ecosystems.

    \item We evaluate the practical usefulness of SBOMs for vulnerability tracking.
    
    \item We publish our replication package\footnote{\url{https://doi.org/10.5281/zenodo.18883005}} to help foment more research on the tooling and SBOM quality topic.
\end{itemize}

We organize this paper across nine remaining sections.
Section~\ref{sec:background} provides key definitions and contextual information related to Software Bill of Materials (SBOMs), laying the foundation for understanding the scope and relevance of our study. 
Section~\ref{sec:methodology} describes our study design in detail, including project selection and SBOM generation. 
Sections~\ref{sec:rq1}, \ref{sec:rq2}, and \ref{sec:rq3} present our findings for each research question, with each section first outlining the specific approach used to address the question and then discussing the corresponding results. 
Section~\ref{sec:discussion} further discusses four aspects of our results, including tool performance, component agreement between tools, and recurring patterns observed across ecosystems. Section~\ref{sec:related_works} reviews the related work, and Section~\ref{sec:limitations} discusses the threats to validity of our study. Finally, Section~\ref{sec:conclusion} concludes the paper.

\section{\textbf{Background}}
\label{sec:background}

\subsection{\textbf{What is an SBOM}}
A Software Bill of Materials (SBOM) is a formal record that contains details and supply chain relationships of the various components used in building software \cite{ntia2020sbomOverview}. 
An SBOM lists all software components, including relevant component metadata, and describes the relationships between them. For example, an SBOM for a web application might list open-source libraries such as React, Lodash, and Axios, including their versions, licenses, and their relationships within the application's dependency hierarchy.
SBOMs offer several key benefits that enhance software development and security processes\cite{sbomsgainingtraction}.
They provide clear visibility into all software components and their dependencies, enabling teams to manage and track both direct and transitive dependencies efficiently. This transparency helps organizations quickly identify and prioritize vulnerabilities, improving incident response and risk management. 
Additionally, SBOMs support license compliance by allowing legal teams to verify that all components adhere to organizational policies. Beyond internal benefits, SBOMs also create a competitive advantage for vendors, as they are increasingly required in procurement processes and signal a commitment to software quality and security \cite{Zahan2023Software}.
\\

\begin{figure}
\centering
\begin{minipage}{0.48\textwidth}
    \centering
    \includegraphics[width=\linewidth]{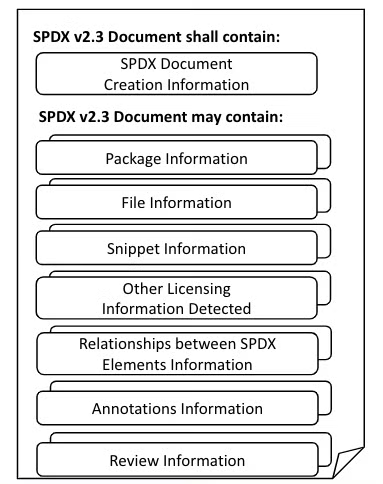}
    \caption{SPDX Document Overview~\cite{SPDX-2.3-overview}}
    \label{fig:spdx-2.3-overview}
\end{minipage}
\hfill
\begin{minipage}{0.48\textwidth}
\begin{lstlisting}[
    basicstyle=\tiny\ttfamily,
    numbers=left,
    numberstyle=\tiny\color{gray},
    frame=single,
    breaklines=true,
    xleftmargin=1.5em,
    framexleftmargin=1em,
    caption={GitHub-generated SBOM Example~\cite{github2025sbomapi}},
    label=lst:github_sbom_example
]
{
  "sbom": {
    "SPDXID": "SPDXRef-DOCUMENT",
    "spdxVersion": "SPDX-2.3",
    "creationInfo": {
      "created": "2021-09-01T00:00:00Z",
      "creators": [
        "Tool: GitHub.com-Dependency-Graph"
      ]
    },
    "name": "github/example",
    "dataLicense": "CC0-1.0",
    "documentNamespace": "https://spdx.org/...",
    "packages": [
      {
        "name": "rails",
        "SPDXID": "SPDXRef-Package",
        "versionInfo": "1.0.0",
        "downloadLocation": "NOASSERTION",
        "filesAnalyzed": false,
        "licenseConcluded": "MIT",
        "licenseDeclared": "MIT",
        "copyrightText": "Copyright (c) 1985",
        "externalRefs": [
          {
            "referenceCategory": "PACKAGE-MANAGER",
            "referenceType": "purl",
            "referenceLocator": "pkg:gem/rails@1.0.0"
          }
        ]
      }
    ],
    "relationships": [
      {
        "relationshipType": "DEPENDS_ON",
        "spdxElementId": "SPDXRef-Repository",
        "relatedSpdxElement": "SPDXRef-Package"
      }
    ]
  }
}
\end{lstlisting}
\end{minipage}
\end{figure}

\noindent
\textbf{The SPDX format.} 
SPDX (System Packet Data Exchange, formerly Software Packet Data Exchange) is an open standard for communicating Software Bill of Materials information, including provenance, licensing, security, and related metadata~\cite{SPDXOverview}. 
By providing a common format, SPDX helps reduce redundant efforts across organizations and communities, facilitating the sharing of critical data and thereby streamlining compliance, security, and reliability processes~\cite{SBOMNTIAFormatSurvey}. 
Figure \ref{fig:spdx-2.3-overview} visually outlines these key attributes, categorizing them into mandatory and optional components in the SPDX v2.3 format.
The core required attribute in any SPDX document is the Creation Information section, which provides the foundational metadata necessary for the document's identification and compatibility with various tools. Additionally, the SPDX format supports several optional sections that allow for a detailed description of the software and its components, including Package Information, File Information, Snippet Information, Other Licensing Information, Relationships, Annotations, and Review Information. 

\noindent
\textbf{Terminology: Components, packages, and dependencies.}
SPDX defines in the \textit{package information} of an SBOM all components that compose the software system. In this paper, we use the term \textit{component} to refer to all components in a software project. However, when an analysis discriminates between the types of dependency components, we refer to them either as \textit{top-level dependencies} or \textit{transitive dependencies} of a software project. 

\subsection{\textbf{GitHub \& SBOM Generation}}

GitHub holds a pivotal role in contemporary software development due to its widespread adoption as a collaborative platform for version control, code sharing, and project management. It serves as a central repository for millions of open source repositories, facilitating efficient teamwork and transparency. While numerous tools exist for generating Software Bill of Materials (SBOMs), the \ghtool stands out for its native integration within the GitHub ecosystem. This integration relies on repositories’ dependency graphs, which summarize manifest and lock files, along with dependencies submitted via the dependency submission API. A dependency graph provides information on dependencies' versions, licenses, manifest sources, and known vulnerabilities, including transitive dependency paths when supported. Importantly, the SBOM generated from this data stays synchronized with the repository’s current dependencies and can be exported in the standard SPDX format via the GitHub UI or REST API \cite{githubSbomExport2025}. This reduces manual work and helps maintain an up-to-date view of a project’s software supply chain. 

Listing~\ref{lst:github_sbom_example} illustrates a \textit{GitHub SBOM} (short for GitHub-generated SBOM) in the SPDX v2.3 JSON format. It shows how the \ghtool captures SPDX metadata such as the SPDX version, creation information, and document namespace. The packages section details a dependency (rails 1.0.0) along with its license (MIT), and package URL. The relationships section explicitly records dependency relationships, indicating which packages the repository depends on.

\subsection{\textbf{NTIA compliance}}

\begin{table}
\centering
\small
\caption{NTIA Minimum Requirements for SBOMs \cite{ntia2021minimum}}
\begin{tabularx}{\textwidth}{l X X}
    \toprule
    \textbf{Data Field} & \textbf{Description} & \textbf{Reason of significance} \\
    \midrule
    Supplier Name & The name of an entity that creates, defines, and identifies components. & It helps trace the distribution source for a specific package as the same component may be supplied through different suppliers (e.g., maven repository, Gradle central repo). \\
    \midrule
    Component Name & Designation assigned to a unit of software defined by the original supplier. & It is crucial to maintain an inventory list of all components used in a software such as packages and files. In GitHub’s generated SBOMs, mainly packages are used. \\
    \midrule
    Version of the Component & Identifier used by the supplier to specify a change in software from a previously identified version. & Versions ensure compatibility and up-to-dateness as it identifies any changes to the software. Therefore, having it is important because we can identify which versions are vulnerable and need updates. \\
    \midrule
    Other Unique Identifiers & Other identifiers that are used to identify a component, or serve as a look-up key for relevant databases. & Identifying components could be confusing so having a unique identifier for each component would make things easier. \\
    \midrule
    Dependency Relationship & Characterizing the relationship that an upstream component X is included in software Y. & Dependency relationships specify different relationship types between components of the software such as which packages depend on others. \\
    \midrule
    Author of SBOM Data & The name of the entity that creates the SBOM data for this component. & Identifies who created the SBOM as there are different tools such as GitHub’s dependency graph. \\
    \midrule
    Timestamp & Record of the date and time of the SBOM data assembly. & Timestamps verify a record of when the SBOM was created for a specific software. It is important because as projects get updated, the SBOM should also change. \\
    \bottomrule
\end{tabularx}

\label{tab:min_req_sbom}
\end{table}

The National Telecommunications and Information Administration (NTIA) has established guidelines for Software Bill of Materials (SBOM) compliance~\cite{nist2022executive,whitehouse2021executive,ntia2021minimum}. 
As detailed in Table~\ref{tab:min_req_sbom}, NTIA compliance requires SBOMs to include seven fundamental data fields that collectively enable comprehensive software component tracking and risk assessment. These minimum elements serve distinct but interconnected purposes in establishing supply chain transparency. The significance of each field, as outlined in the table, demonstrates how NTIA's requirements address critical aspects of software security management, from tracing distribution sources and maintaining component inventories to identifying vulnerable versions and understanding inter-component dependencies. Additionally, compliant SBOMs must support automation and be machine-readable, typically delivered in standardized formats such as SPDX, CycloneDX, or SWID tags \cite{ntia2019_framing}. This regulatory framework represents a significant shift toward mandating supply chain transparency, particularly for software vendors serving federal government clients, while establishing a foundation for broader industry adoption of systematic software component tracking and risk assessment practices that leverage the structured data fields specified in NTIA's minimum requirements.

\section{\textbf{Methodology}}
\label{sec:methodology}

\begin{figure}[ht]
    \centering
    \includegraphics[width=0.98\linewidth]{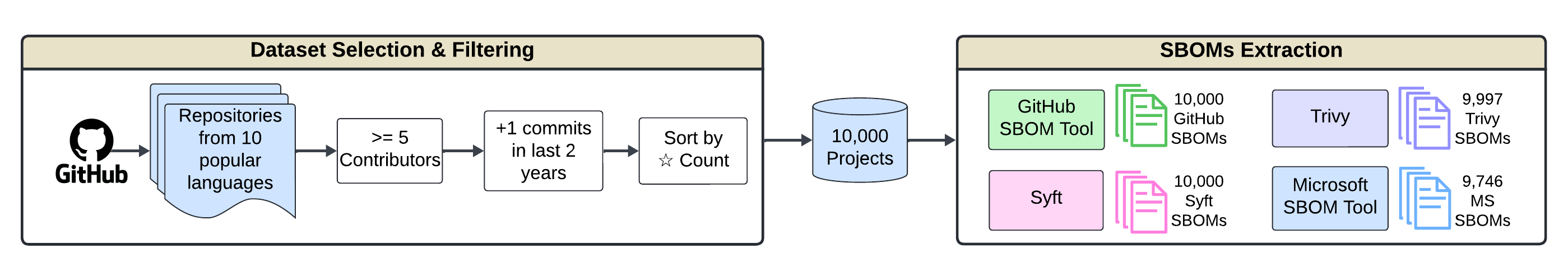}
    \vspace{-0.2pt}
    \includegraphics[width=1\linewidth]{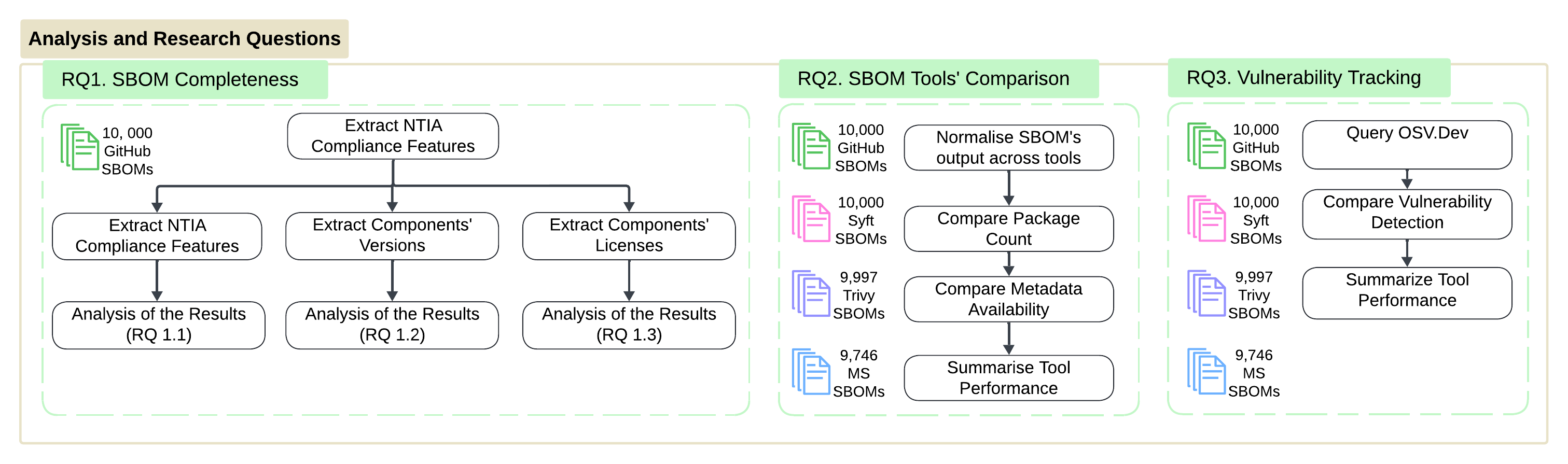}
    \caption{Methodology overview of this study. }
    \label{fig:stacked}
\end{figure}

Our goal is to study the completeness and consistency of SBOMs generated by GitHub across 10 programming ecosystems. 
Figure~\ref{fig:stacked} illustrates our methodology. 
We begin by selecting 1,000 projects from each programming language, aiming to capture active, highly popular, and collaborative software repositories (Section~\ref{subsec:project_selection}). 
For each repository, we generate a GitHub SBOM, yielding a total of 10,000 SBOMs (Section~\ref{subsec:sbom_extraction}). To enable comparative analysis, we also generate SBOMs for the same repositories using three widely used tools: Syft, Trivy, and the \mstool (Section~\ref{subsec:sbom_extraction}). We evaluate GitHub SBOMs for NTIA minimum compliance, version coverage, and license availability (RQ1, Section~\ref{sec:rq1}). We then compare GitHub SBOMs with SBOMs generated by other tools to assess differences in metadata completeness and dependency detection (RQ2, Section~\ref{sec:rq2}). Finally, we examine the practical usefulness of SBOMs for detecting known vulnerabilities (RQ3, Section~\ref{sec:rq3}). Together, these analyses provide a comprehensive view of SBOM completeness, metadata quality, and practical utility across programming languages and tooling.

\subsection{Project Selection}
\label{subsec:project_selection}

To analyze SBOMs generated by GitHub, we target \emph{open-source} repositories that are \emph{widely used} and \emph{actively maintained} by multiple contributors. 
These projects are more likely to reflect mature development practices and realistic software supply-chain behaviours. This choice is motivated by prior evidence that GitHub contains a substantial amount of repositories that are personal or inactive, and that naive sampling can therefore lead to datasets that are not representative of collaborative software development \cite{Kalliamvakou2014PromisesPerils}. 

Within this population, we strategically select projects from ten widely used programming languages: C, C++, C\#, Java, PHP, Python, JavaScript, Go, Rust, and Swift. These languages were chosen because they are consistently reported among the most popular and widely adopted languages across complementary popularity indicators, including the TIOBE Programming Community Index \cite{TIOBEIndexDec2025}, GitHub’s Octoverse language rankings \cite{GitHubOctoverse2024}, and large-scale developer surveys from Stack Overflow \cite{StackOverflowSurvey2025Technology}. 

\textbf{Step 1: Contributor Threshold.} We select only repositories with at least \textbf{five unique contributors}. This criterion helps us focus on \emph{collaborative} and \emph{community-driven} projects (instead of personal/toy repositories), which are more likely to reflect real-world development and maintenance practices. Similar contributor-based thresholds are commonly used in prior empirical studies in software engineering~\cite{papamichail2019roles,jafari2022depsmells}.

\textbf{Step 2: Recent Maintenance.} The software repository must have shown at least one commit within the two years prior to our data extraction. This condition ensures that selected projects are minimally maintained.
For the filtering in Steps~1 and~2, we use the SEART GitHub Search (GHS) platform~\cite{SEARTGHS}, a web-based tool designed to support reproducible sampling of GitHub repositories for mining software repositories studies. SEART-GHS maintains a continuously updated dataset of GitHub repository metadata and provides a query interface that enables filtering projects based on multiple criteria, such as programming language, repository activity, popularity indicators, and licensing information ~\cite{Dabic2021Sampling}. SEART-GHS has been adopted in prior empirical software engineering studies to select representative sets of actively developed repositories~\cite{9978190}~\cite{Mastropaolo2024ICPC}~\cite{Romeo2025UML}, demonstrating its suitability for constructing reliable and reproducible datasets.

\begin{figure}[ht]
    \centering
    \includegraphics[scale=0.5]{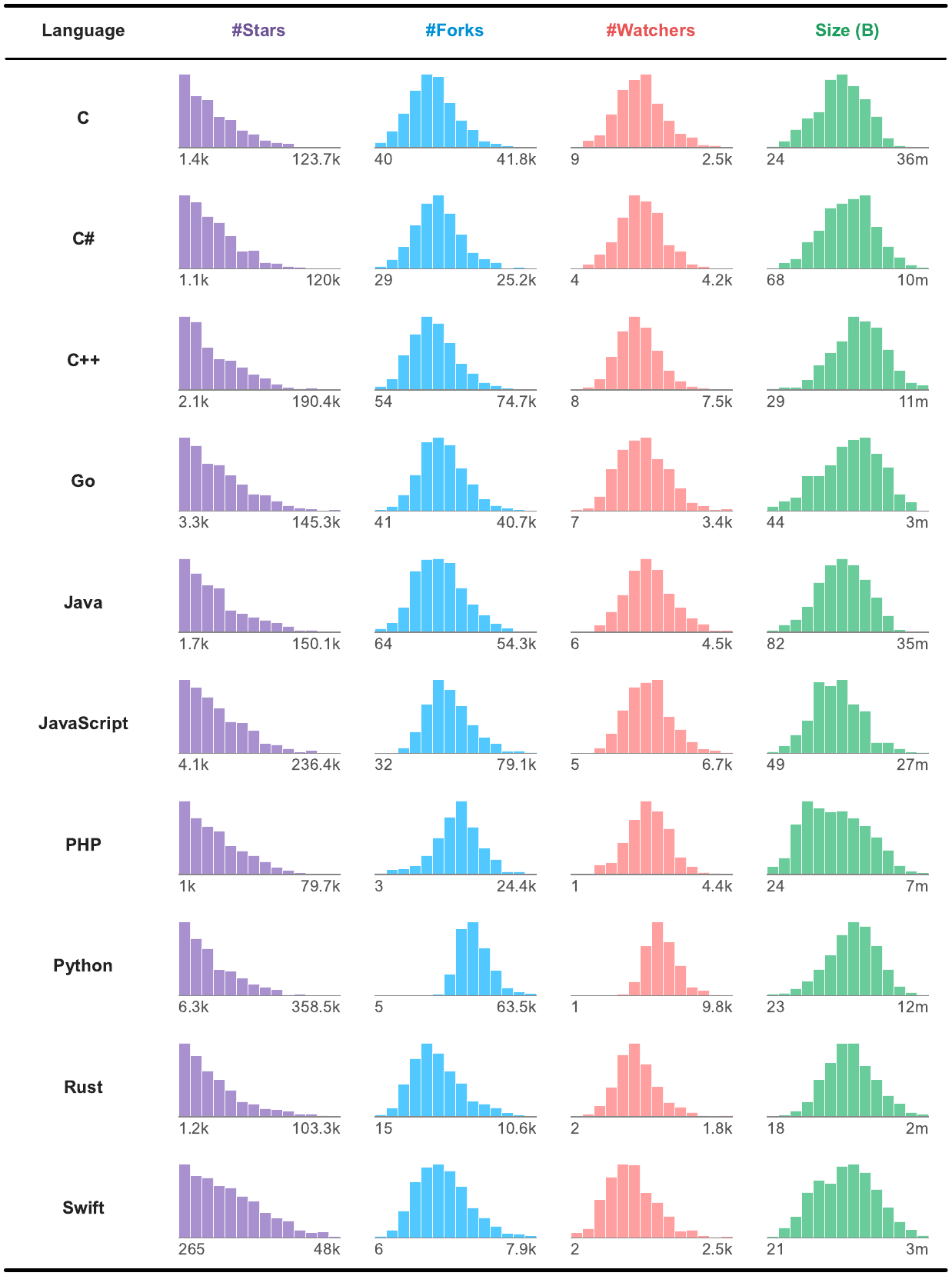}
    \caption{Statistical Distribution of the Selected 10,000 Repositories}
    \label{fig:repo_stats_visualization}
\end{figure}

\textbf{Step 3: Popularity by Stars.} We select the \textbf{top 1,000 most popular projects per programming language} based on their GitHub star count, resulting in a total of \textbf{10,000 projects across 10 programming languages}. GitHub allows users to \emph{star} repositories to express interest in or appreciation of a project. Prior empirical research has shown that the number of stars is commonly used as a proxy for project popularity, where repositories with higher star counts tend to receive greater community attention and adoption \cite{Borges2016Popularity}. 
Following this established practice, we prioritize projects with higher star counts to capture influential and widely adopted software systems. This focus on popular repositories increases the likelihood that the analyzed projects reflect mature development practices and have broader ecosystem impact, thereby improving the generalizability and relevance of our findings.
Following these steps, we selected 1,000 repositories for each of the 10 programming languages, resulting in a total dataset of 10,000 repositories. 
We present the distribution of the selected projects across different characteristics in Figure~\ref{fig:repo_stats_visualization}.

\subsection{Selection of SBOM Generation Tools}

To compare the quality of GitHub SBOMs, we select the other widely used SBOM generation tools:

\begin{itemize}
    \item \textbf{Syft:} An open-source CLI tool developed by Anchore that generates SBOMs from container images, filesystems, and source directories, supporting multiple formats including SPDX and CycloneDX~\cite{syft2025}.
    
    \item \textbf{Trivy:} A security and compliance scanner maintained by Aqua Security that can produce SBOMs as well as detect vulnerabilities in container images, filesystems, and Git repositories~\cite{trivy2025}.
    
    \item \textbf{\mstool:} A free command-line tool provided by Microsoft that generates SBOMs for a wide range of project types, supporting multiple programming languages and producing output in SPDX and CycloneDX formats~\cite{microsoftSbomToolHome2025}.

\end{itemize}

We selected these tools because they are free, open‑source, and support multiple operating systems (Linux, Windows, and macOS). These tools are also frequently used in previous studies ~\cite{Yu2024CorrectnessSBOM}~\cite{Torres-Arias2023A} and industry overviews ~\cite{oxSecuritySbomTools2025}~\cite{wiz_top_sbom_tools_2025}, demonstrating their relevance and adoption.

\subsection{SBOM Generation}
\label{subsec:sbom_extraction}

Upon selecting our dataset of the most popular 10,000 open-source repositories across 10 programming languages, we aim to export their SBOMs using the GitHub SBOM generator, Trivy, Syft, and \mstool.   

\noindent
\textbf{\ghtool.}
To extract SBOMs generated from the recently added \ghtool, we use the GitHub REST API for SBOMs ~\cite{github2025sbomapi} to programmatically generate the SBOM for each selected repository by issuing a \texttt{GET} request to the \texttt{/repos/{owner}/{repo}/dependency-graph/sbom} endpoint, which returns an SPDX-compatible SBOM derived from GitHub’s dependency graph when enabled.
All of the SBOMs were obtained in SPDX JSON format, as GitHub's SBOM REST API currently exports SBOMs in this format ~\cite{github2025sbomapi}.
For each language, we initiated SBOM extraction starting from the most popular repositories and proceeded down the list. Since GitHub requires the dependency graph to be enabled in a repository to export an SBOM ~\cite{githubSbomExport2025}, we received a 404 error when attempting to generate SBOMs via the REST API for repositories where the dependency graph was disabled by the owners. In such cases, we skipped the repository and continued to the next most popular one. This iterative process was repeated until we successfully obtained 1,000 valid SBOMs per language.
\\

\noindent
\textbf{Extraction from Local Repositories.}
All three SBOM generator tools, Trivy, Syft, and \mstool, provide support for SBOM generation using local repositories~\cite{trivyRepository2025,syftSupportedSources2025,microsoftSbomTool2025}.
Following the GitHub-based SBOM generation, all 10{,}000 repositories were cloned to local storage immediately thereafter, and all local SBOM generation runs were performed against these same cloned snapshots. By cloning directly after the GitHub phase, the local snapshots reflect the same repository state that was present during GitHub SBOM generation, minimizing the risk of source divergence between the two extraction methods. Each tool was then executed against the same locally cloned repository snapshot, ensuring identical source code, dependency manifests, and project structures were analyzed across all three tools.
Each tool has parameters that control the sensitivity of the SBOM generation.
We configured each tool to run under the most permissive license-discovery settings available to obtain the maximum possible license coverage

\medskip
\noindent
\textbf{Trivy.}
For Trivy, SBOMs were generated from the locally cloned repositories using Trivy’s filesystem scanning mode. For each repository, we executed Trivy with SPDX JSON output enabled. 
We enabled \texttt{--license-full}, which instructs it to scan not only component metadata but also the full contents of each file for embedded license expressions.
Trivy successfully generated SBOMs for 9{,}997 repositories. Three failures occurred due to excessive analysis time on large multi-module Java projects and a stack-overflow issue triggered by its XML-based Go module parsing.

\medskip
\noindent
\textbf{Syft.}
Syft operates exclusively on local inputs such as source code directories, container images, or packaged artifacts~\cite{syftSupportedSources2025}. For each cloned repository, Syft was executed with the repository directory as input and configured to emit SPDX JSON output. Syft successfully generated SBOMs for all 10{,}000 repositories in the dataset.

\medskip
\noindent
\textbf{\mstool.}
The \mstool similarly requires local repository inputs~\cite{microsoftSbomTool2025}. For each repository, SBOM generation involved creating a temporary manifest directory and executing the tool with repository metadata parameters. 
The \mstool internally queries the ClearlyDefined ~\cite{clearlydefined} API, an open-source crowdsourced database of license and attribution metadata, to retrieve license information.
The resulting SPDX JSON files were extracted from the generated manifest directory. The \mstool exhibited a higher failure rate, particularly on repositories containing recursive symbolic-link structures or missing build-generated files. These failures stem from limitations in its hashing mechanism, which cannot correctly handle symlink loops or absent files, resulting in a total of 9{,}746 successfully generated SBOMs.
\\

\section{RQ1: \rqone}
\label{sec:rq1}
Understanding the completeness of GitHub-generated SBOMs is essential because the usefulness of any SBOM depends on the accuracy and availability of the information it provides~\cite{ntia2019_framing}. 
While specific needs may vary by use case, all applications require a consistent and systematic approach to defining and identifying software components and their relationships. An SBOM typically includes both the primary software component and its dependency components. 
In this study, we focus on the latter. 

We assess the completeness of the generated SBOMs with respect to three main aspects: (1) the NTIA minimum requirements (see Table~\ref{tab:min_req_sbom}), (2) the presence of version information for dependencies, and (3) the availability of licensing information for dependencies. Accordingly, we structure \textbf{RQ1} into three sub-questions:

\begin{itemize}
    \item \textbf{RQ 1.1:} Do GitHub SBOMs satisfy the NTIA minimum requirements?
    \item \textbf{RQ 1.2:} To what extent do GitHub SBOMs include \textbf{version information} for dependencies?
    \item \textbf{RQ 1.3:} To what extent do GitHub SBOMs provide \textbf{licensing information} for dependencies?
\end{itemize}

\subsection{Approach}

To evaluate the completeness of GitHub SBOMs, we analyze each SBOM 
and aggregate the results by programming language to identify ecosystem-level patterns.
\\

\noindent\textbf{Assessing NTIA Compliance of GitHub-Generated SBOMs.}
According to the NTIA guidance on the Minimum Elements for an SBOM, an SBOM should list all primary components along with their dependencies~\cite{ntia2021minimum}. 
We analyze each SBOM file to identify the NTIA-required elements (listed in Table~\ref{tab:min_req_sbom}) using the corresponding SPDX field names and specification references.

\begin{itemize}
    \item \textbf{Component Name:}  
    The primary component, i.e., the software system for which the SBOM was generated, is identified from the top-level \texttt{name} field of the SBOM document. Dependency components' names are found by inspecting the \texttt{name} field within each entry of the \texttt{packages} array.

    \item \textbf{Version of the Component:}  
    The version of each component—including the primary component and third-party library packages—is extracted from the \texttt{versionInfo} field located in its respective entry within the \texttt{pack-\\ages} array. 

    \item \textbf{Supplier Name:}  
    The supplier is extracted by checking the \texttt{supplier} or \texttt{PackageSupplier} field within each package entry.

    \item \textbf{Other Unique Identifiers:}  
    Unique identifiers for components are found in the SPDXID field, while the SBOM document itself is uniquely referenced through the \texttt{documentNamespace} field.
    Examples of commonly used unique identifiers are Common Platform Enumeration (CPE), Software Identification (SWID) tags, and Package Uniform Resource Locators (PURL) \cite{ntia2021minimum}. 
    We tried to look for CPE, SWID or PURL by inspecting the \texttt{externalRefs} field of each entry of the \texttt{packages} array.
    We examine the \texttt{externalRefs} field of each entry in the \texttt{packages} array and consider the presence of a unique identifier if at least one of CPE, SWID, or PURL is found.

    \item \textbf{Dependency Relationship:}  
    When present, dependency relationships between components are described in the \texttt{relationships} array.

    \item \textbf{Author of SBOM Data:}  
    The authoring entity is identified through the \texttt{creators} field under the \texttt{creationInfo} section. This may include the names of tools, individuals, or organizations responsible for generating the SBOM.

    \item \textbf{Timestamp:}  
    The timestamp indicating when the SBOM was created is retrieved from the \texttt{created} field within the \texttt{creationInfo} section.
\end{itemize}
\noindent

NTIA recommends that SBOMs include all top-level dependencies with sufficient detail to enable the recursive discovery of transitive dependencies~\cite{ntia2021minimum}.
Therefore, in this analysis, we focus primarily on the top-level dependencies for each project and verify whether each of them includes required fields such as \texttt{name}, \texttt{supplier}, \texttt{versionInfo}, and a unique identifier.
We determine the top-level dependencies by examining the \texttt{relationships} array and selecting entries where the \texttt{spdxElementId} matches this primary identifier.
\\

\noindent\textbf{Comparison of Version Coverage of Top-Level and Transitive Dependencies.} 
Beyond just the NTIA compliance, which considers just the top-level dependencies, it is also beneficial to include version information for transitive dependencies. 
Transitive dependencies are known to introduce security vulnerabilities, licensing risks, or compatibility issues~\cite{xu2023remediating}\cite{mir2023maven}.
To identify transitive dependencies, we trace the dependencies from each top-level dependency to the components it depends on. 
After identifying both primary and transitive dependencies, we inspect the \texttt{versionInfo} field in their corresponding entries within the \texttt{packages} array to verify whether valid version information is provided. 
\\

\noindent\textbf{Identifying Components' License Availability.}
For each SBOM, we checked whether license information was available for the top-level dependencies and transitive dependencies. Within the \texttt{packages} array, the \texttt{licenseConcluded} and \texttt{licenseDeclared} fields were checked for all the dependencies. 
We measured the share of components with available license information among top-level dependencies, transitive dependencies, and all dependencies combined (the union of both). These dependency-level percentages were then averaged by language to determine overall license availability trends.

\subsection{Do GitHub-generated SBOMs satisfy the NTIA minimum requirements?}

We present the NTIA compliance results for all analyzed SBOMs, summarized in Table~\ref{tab:ntia_compliance_by_lang}, including a breakdown by element, attribute, and programming language. 
We observe the following:

\textbf{Finding 1. No GitHub SBOM is fully compliant with the NTIA minimum requirements.}
To be considered fully compliant, an SBOM must satisfy all defined minimum elements, and in our analysis, that was obtained by no SBOM generated by GitHub.
The most critical gap is the absence of supplier information---0\% of the SBOMs include supplier data for either the primary component or any top-level dependencies. Supplier information is an important information as it specifies the name of the entity that creates, defines and identifies components.
For example, \textit{Guava} is a Java library developed and distributed by \textit{Google}; an SBOM that lists \textit{Guava} as a dependency should therefore identify \textit{Google} as its supplier. This universal omission of supplier data leads to 0\% full compliance with the NTIA minimum requirements across all evaluated languages.

\textbf{Finding 2. GitHub SBOMS show strong compliance in core metadata fields.}
Despite the absence of supplier information, all other NTIA-required fields are widely represented and compliant across SBOMs.
Every SBOM includes the name and version of the primary component, as well as their unique identifier. In addition, 100\% of SBOMs provide dependency relationships for all listed components, author metadata, and timestamps.
Additionally, all SBOMs include the name and unique identifiers for all the top-level dependencies. 
For GitHub SBOMs, all the top-level dependencies include SPDXID as well as PURL as their unique identifier. 

\textbf{Finding 3. Dependency version coverage in projects varies significantly per programming language.}
The level of compliance varies substantially across projects written in different programming languages.
Swift exhibits the highest coverage, with 98.8\% of SBOMs providing version information for all top-level dependencies, followed closely by JavaScript (96.2\%), Go (93.5\%), and Rust (91.8\%). In contrast, Java and Python show the lowest compliance in this area, with respectively only 41.4\% and 29.9\% of their SBOMs including version information for all the top-level dependencies. C, C\#, C++, and PHP fall in the mid-to-high range, with coverage between 75.0\% and 83.3\%.

\begin{table}
\centering
\small
\setlength{\tabcolsep}{2.5pt}
\renewcommand{\arraystretch}{1.15}
\caption{Compliance of SBOMs Across Different Languages with NTIA's Minimum Requirements}
\resizebox{\textwidth}{!}{%
\begin{tabular}{p{2.4cm}p{1.8cm}*{10}{r}}
\toprule
\textbf{Element} & \textbf{Attribute} & \textbf{C} & \textbf{C\#} & \textbf{C++} & \textbf{Go} & \textbf{Java} & \textbf{JS} & \textbf{PHP} & \textbf{Python} & \textbf{Rust} & \textbf{Swift} \\
\midrule
\multirow{4}{*}{\parbox{2.8cm}{\raggedright Primary Software Component}} 
  & Supplier & 0\% & 0\% & 0\% & 0\% & 0\% & 0\% & 0\% & 0\% & 0\% & 0\% \\
  & Name & 100\% & 100\% & 100\% & 100\% & 100\% & 100\% & 100\% & 100\% & 100\% & 100\% \\
  & Version & 100\% & 100\% & 100\% & 100\% & 100\% & 100\% & 100\% & 100\% & 100\% & 100\% \\
  & Unique ID & 100\% & 100\% & 100\% & 100\% & 100\% & 100\% & 100\% & 100\% & 100\% & 100\% \\
\midrule
\multirow{4}{*}{\parbox{2.8cm}{\raggedright All Top-Level Dependency Components}} 
  & Supplier & 0\% & 0\% & 0\% & 0\% & 0\% & 0\% & 0\% & 0\% & 0\% & 0\% \\
  & Name & 100\% & 100\% & 100\% & 100\% & 100\% & 100\% & 100\% & 100\% & 100\% & 100\% \\
  & Version & 83.3\% & 79.5\% & 75.0\% & 93.5\% & 41.4\% & 96.2\% & 82.9\% & 29.9\% & 91.8\% & 98.8\% \\
  & Unique ID & 100\% & 100\% & 100\% & 100\% & 100\% & 100\% & 100\% & 100\% & 100\% & 100\% \\
\midrule
\multirow{1}{*}{\raggedright All Components} 
  & \raggedright Dependency Relationships & 100\% & 100\% & 100\% & 100\% & 100\% & 100\% & 100\% & 100\% & 100\% & 100\% \\
\midrule
\multirow{2}{*}{\raggedright SBOM} 
  & Author & 100\% & 100\% & 100\% & 100\% & 100\% & 100\% & 100\% & 100\% & 100\% & 100\% \\
  & Timestamp & 100\% & 100\% & 100\% & 100\% & 100\% & 100\% & 100\% & 100\% & 100\% & 100\% \\
\midrule
\textbf{\raggedright Overall\\ Compliance} &  & 0\% & 0\% & 0\% & 0\% & 0\% & 0\% & 0\% & 0\% & 0\% & 0\% \\
\bottomrule
\end{tabular}%
}

\label{tab:ntia_compliance_by_lang}
\end{table}

\subsection{Do GitHub-generated SBOMs include version information for dependencies?}

\begin{figure}
    \centering
    \includegraphics[width=1\linewidth]{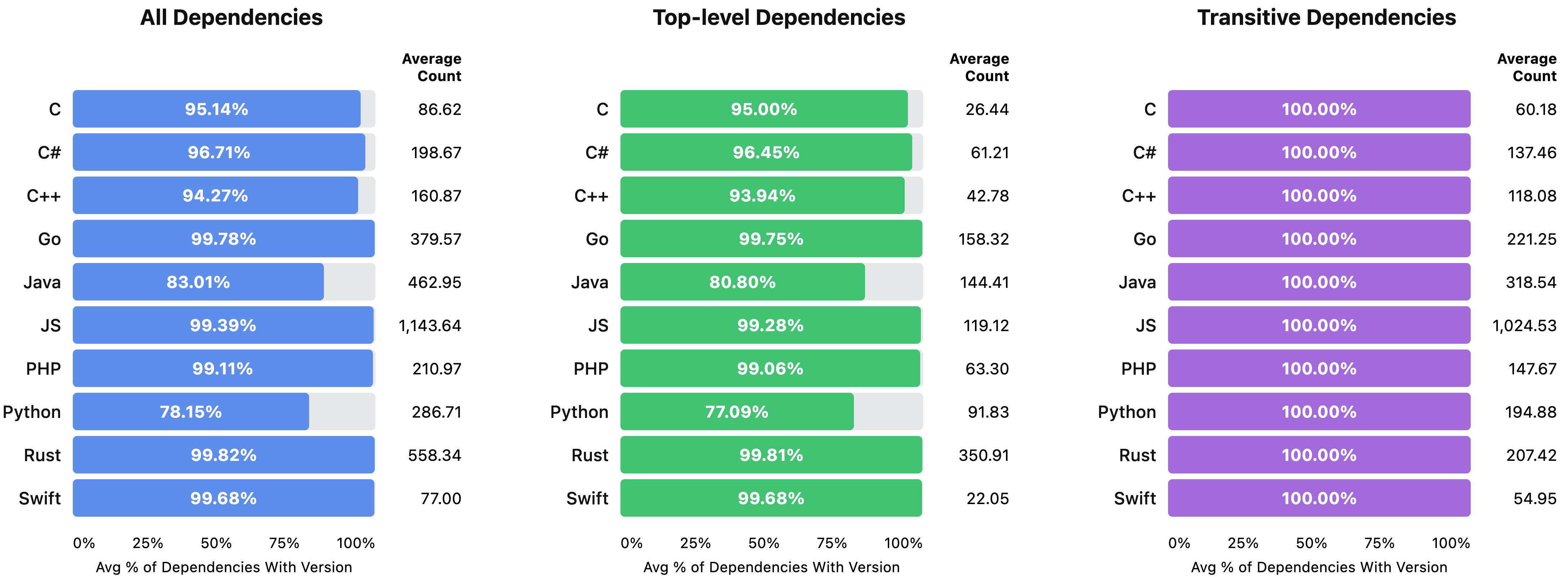}
    \caption{Average version information of dependencies per project.} 
    \label{fig:avg_count_and_versioned_percentage_of_deps_by_lang}
\end{figure}

We present the results of our analysis in Figure~\ref{fig:avg_count_and_versioned_percentage_of_deps_by_lang}. 
We observe significant variation in the consistency with which generated SBOMs contain dependency version information across languages.
It is important to note that the values in the table differ significantly from those in our analysis in RQ1. RQ1 focuses on NTIA compliance and shows the share of projects that include version information for all top-level dependencies, while in this analysis, we are interested in the average distribution of version information across projects. 

\textbf{Finding 1. With the exception of Python and Java projects, GitHub-generated SBOMs, on average, include more than 94\% of versioned dependencies.} 
We show in Figure~\ref{fig:avg_count_and_versioned_percentage_of_deps_by_lang} that the presence of versioned dependencies encompasses at least 94\% of a project's dependencies across most of the evaluated programming languages. 
The share of versioned dependencies exceeds 99.5\% in projects in Go, Rust, and Swift, showing the strongest support across the ten languages. 

The exception to this rule is projects in Java and Python. 
Java projects exhibit an average of 83\% of versioned dependencies, while Python projects show only 78\% of versioned dependencies. 
These results are inline with our results in RQ1 (Table~\ref{tab:ntia_compliance_by_lang}), where we showed that Java and Python projects exhibited the least share of projects with version included in all top-level dependencies. 
C, C++, and C\# projects fall between these extremes, with approximately 95\% of dependencies on average specifying version information. 

\textbf{Finding 2. All listed transitive dependencies have version information, but transitive dependencies of unversioned dependencies are omitted.} 
Figure~\ref{fig:avg_count_and_versioned_percentage_of_deps_by_lang} also breaks down the analysis per top-level dependency (a.k.a. direct dependencies) and transitive dependencies. 
We notice that all transitive dependencies include version information across all programming languages. 
This apparent completeness arises from the behavior of the \ghtool.
We examined the generated SBOMs in depth and found that when a parent dependency lacks version information, the tool omits all of its transitive dependencies from the SBOM. 

In these cases, only the unversioned parent dependency appears, with none of its downstream dependencies listed. 
Consequently, the reported 100\% version coverage for transitive dependencies does not imply that all transitive dependencies in the projects are versioned; it reflects a conservative approach from the \ghtool, that only includes transitive dependencies of versioned parent dependencies.

\textbf{Finding 3. There are vast differences in the number of project components across programming languages.}
Similarly, as reported in previous studies~\cite{Decan2019}, the number of dependencies varies substantially across programming languages. 
On average, JavaScript and Rust projects exhibit the highest total dependency counts, with 1143 and 558 dependencies per project, respectively, indicating a strong reliance on third-party libraries. In contrast, Swift and C projects show considerably lower dependency usage, averaging 77.00 and 86.62 dependencies per project.

According to GitHub SBOMs, we observe distinct patterns in how programming languages structure and manage their dependency hierarchies: some ecosystems exhibit broad, direct composition, whereas others rely on deeper chains of indirect dependencies. In particular, Rust, Go, and Java projects tend to include more top-level dependencies on average than other languages, suggesting heavier reliance on direct imports. However, this pattern reverses at the transitive level. On average, JavaScript projects include over 1,024 transitive dependencies, approaching an order-of-magnitude increase relative to their direct imports, and substantially exceeding the transitive dependency counts observed in Rust, Go, and Java projects. This contrast highlights JavaScript projects' exceptionally deep dependency trees, where relatively fewer direct components expand into extensive chains of indirect reuse. Python, PHP, C, C++, C\#, and Swift projects exhibit moderate expansion, typically with two to three times more transitive dependencies than top-level components, reflecting layered but more tightly managed dependency structures.

\begin{figure}
    \centering
    \includegraphics[scale=0.5]{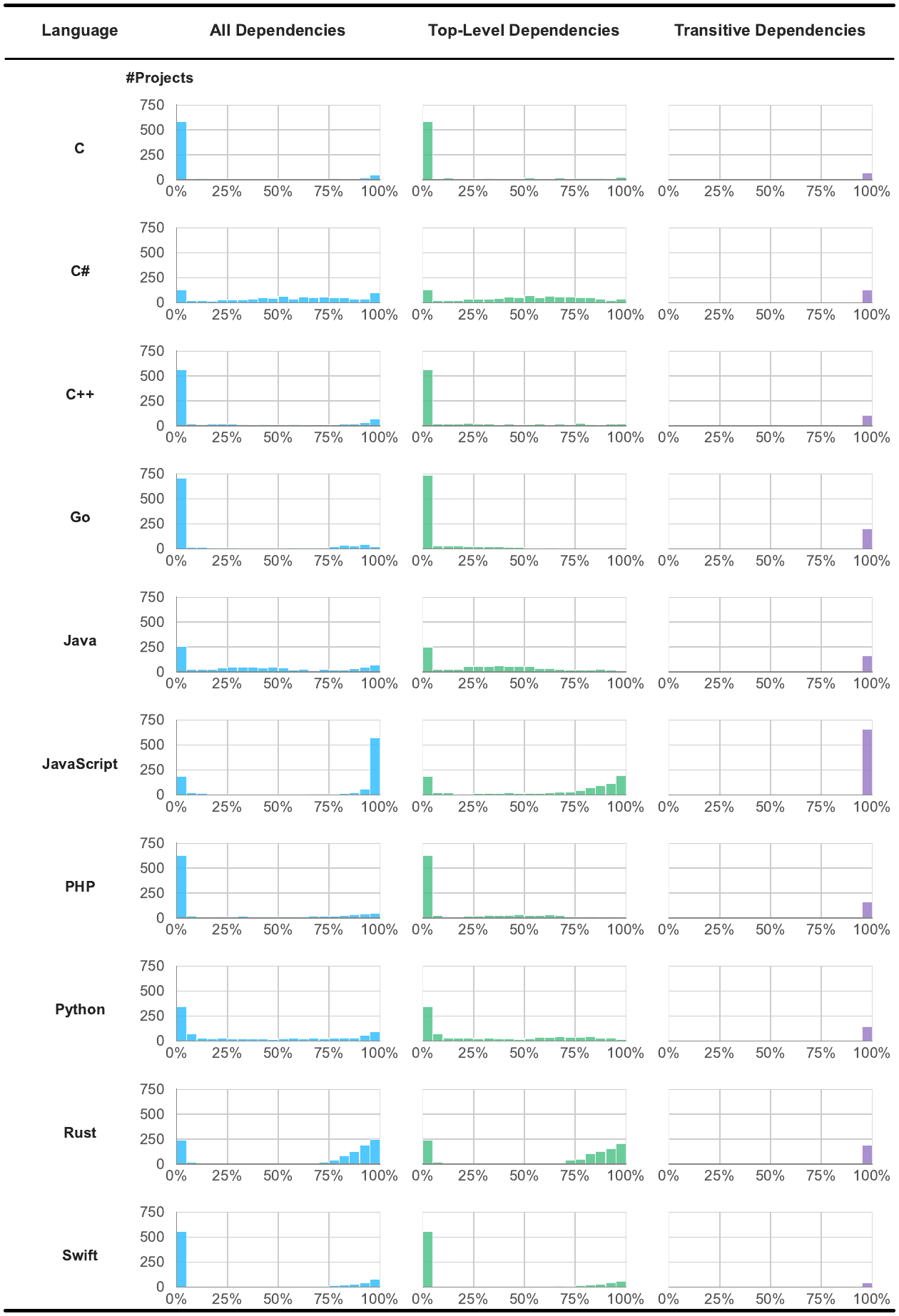}
    \caption{The distribution of dependencies' license information per project. Unlike top-level dependencies, not all generated SBOMs include transitive dependencies; thus, the count may not sum up to 1,000 per programming language.  
    }
    \label{fig:licensed_percentage_of_deps_by_lang}
\end{figure}

\subsection{Do GitHub-generated SBOMs provide \textbf{licensing information} for dependencies?}
We present a distribution of the percentage of dependencies with licensing information per project, broken down by programming language and dependency type (all, top-level, and transitive) in Figure ~\ref{fig:licensed_percentage_of_deps_by_lang}. 
We opt to showcase the distributions rather than averages because, unlike the component version analysis, the data is heavily skewed. 
Many projects have either 0\% or 100\% of dependencies licensed, making the mean a poor summary of the underlying variation. 
Once again, we notice some contrasts related to licensing information across programming languages, and report the following observations: \\

\textbf{Finding 1. For projects written in C, C++, Go, PHP, Python, and Swift, license information is absent for the vast majority of dependencies.} 
Figure~\ref{fig:licensed_percentage_of_deps_by_lang} shows that, for C, C++, Go, PHP, and Swift projects, the distribution of license coverage is close to bimodal. More than half of the projects in each of these languages report 0\% of their dependencies with license information, while only a small fraction achieve complete (100\%) coverage. Very few projects fall in the intermediate ranges.

\textbf{Finding 2. SBOMs from JavaScript and Rust projects report license information for a substantially larger share of dependencies.}
In contrast to other ecosystems, JavaScript and Rust projects demonstrate comparatively high levels of license information reporting. In JavaScript, more than half of the projects provide license information for nearly all dependencies. 
Rust projects exhibit a similar trend, with nearly one-quarter achieving complete or near-complete license reporting. This observation is consistent with the findings of Wu et al.~\cite{wu2024_license}, who analyzed 33,710,877 packages across five package managers and reported that npm (JavaScript) and Cargo (Rust) exhibit the highest percentages license information in SBOMs among all studied ecosystems. 

\textbf{Finding 3. C\#, Java, and Python projects have varying degrees of license information.}
When analyzing projects written in C\#, Java, and Python, we notice a more uniform distribution in the proportion of dependencies for which license information is reported. 
GitHub-generated SBOMs vary in the degree of license information; however, across all ecosystems, fewer than 25\% of projects provide license information for all their dependencies. 
This suggests that most projects will have a sizeable share of components without license information in the GitHub-generated SBOMs for these programming languages.

\textbf{Finding 4. License information is almost always present in listed transitive dependencies.}
The differences in license availability across ecosystems are mainly due to how the \ghtool reports licenses for top-level dependencies, as shown in Fig.~\ref{fig:licensed_percentage_of_deps_by_lang}. 
However, for the listed transitive dependencies, license information is almost always included, regardless of the language. 
As discussed in our previous analysis, \ghtool adopts a conservative approach and only lists transitive dependencies of parent dependencies that have a version. 

\section{RQ2: \rqtwo}
\label{sec:rq2}
While RQ1 focused exclusively on evaluating the completeness of SBOMs generated by GitHub, this research question broadens the scope to assess how GitHub’s SBOM tool compares with other widely used SBOM generation tools in practice. 
The goal is to understand whether GitHub SBOMs provide comparable metadata quality, particularly in attributes required for effective vulnerability analysis and license compliance assessment.
Accordingly, we analyze SBOMs generated by GitHub, Trivy, Syft, and the \mstool for the selected repositories. The procedures used to generate these SBOMs are described in Section~\ref{sec:methodology}; in this research question, we focus exclusively on analyzing and comparing the contents of the resulting SBOMs.

\subsection{Approach}
\label{subsec:rq2_approach}

Using the SBOMs generated for the selected repositories, we analyze differences at two levels: dependency detection and metadata completeness. We first examine variations in the number of components reported by each tool to capture differences in dependency discovery. We then assess the presence of version and license metadata at the component level, applying consistent validation criteria and aggregating results at the repository level to enable comparison across tools and programming languages.
\\

\noindent\textbf{Component Count Analysis.} For all the SBOMs generated by the four tools, we first examined the number of components reported in each SBOM. 
Due to the highly skewed distribution of component count data across SBOM tools, we employ two statistical methods, as done in previous studies~\cite{frankford2026} \cite{liu2021technicaldebt}.
First, we employed the Kruskal–Wallis H-test \cite{KruskalWallis1952}, a non-parametric test used for comparing multiple independent samples. 
For each programming language, this test was used to assess whether the four SBOM tools differ statistically in the number of detected components across repositories. When the Kruskal–Wallis test indicated significant differences, we conducted Dunn's post-hoc test \cite{Dunn:1961:MCA} with Bonferroni correction to examine pairwise differences in component count distributions between the \ghtool and each of the other SBOM generation tools. 
\\

\noindent\textbf{Validation Criteria.}
We choose to focus on version and license information to assess the comparative performance of the four SBOM tools.
Each metadata attribute was evaluated for presence and validity using the following criteria:

\begin{itemize}
    \item \textbf{Version information} was considered present if an SBOM entry contained a non-empty \texttt{versionInfo} or \texttt{version} field that was not marked as ``UNKNOWN''.

    \item \textbf{License information} was considered present if any of \texttt{licenseDeclared},\\ \texttt{licenseConcluded}, or \texttt{licenseInfoFromFiles} contained meaningful values other than ``NOASSERTION'', ``NONE'', or ``UNKNOWN''.
\end{itemize}

\noindent\textbf{Percentage Calculation and Aggregation.}
For each SBOM, we computed the percentage of components that contained valid values for each metadata attribute. For every tool--language combination, these per-SBOM percentages were averaged across all valid SBOM files. 
This repository-centric aggregation ensures each repository contributes equally to the final results, regardless of its dependency graph size.

\subsection{Results}
\label{subsec:rq2_findings}
We present here the main empirical findings from our analysis, highlighting differences in component detection and metadata reporting across SBOM tools and programming languages.\\

\begin{figure}
    \centering
    \includegraphics[width=1\linewidth]{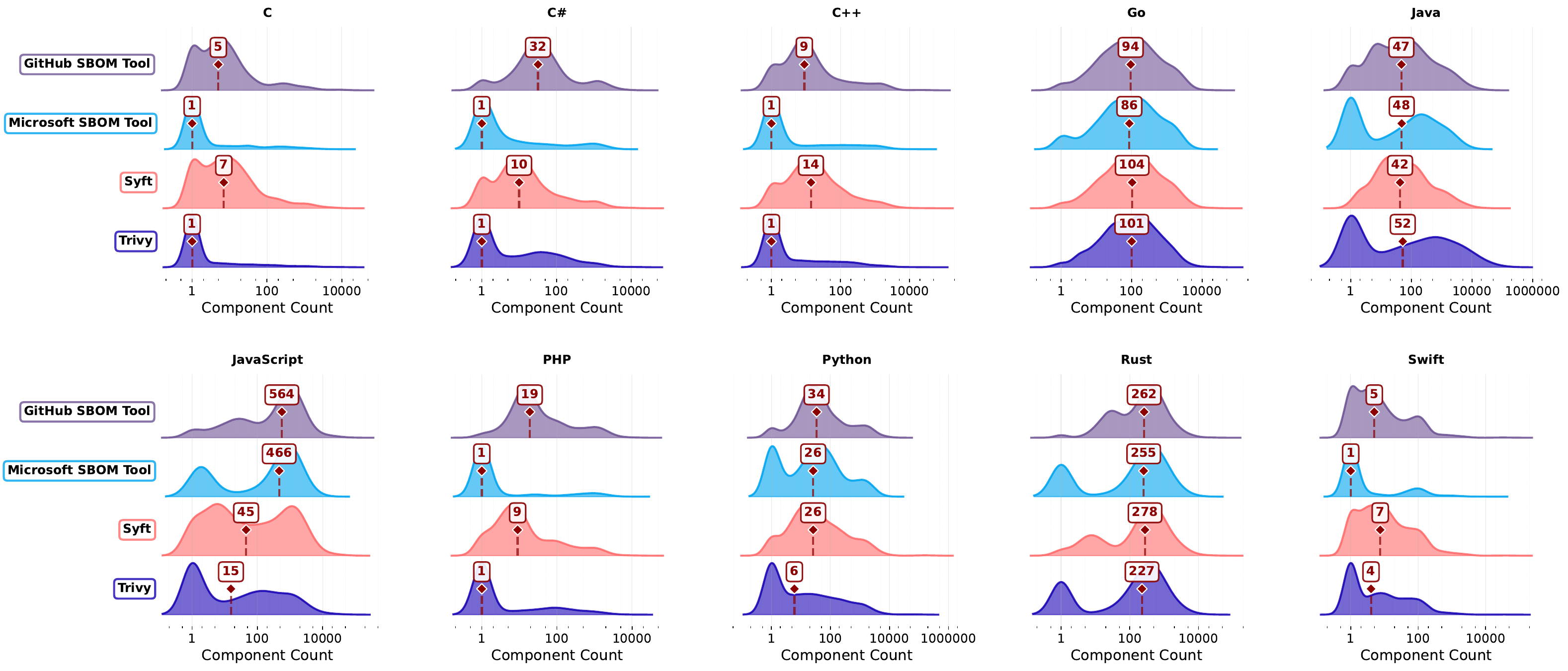}
    \caption{Component Count Distribution Per Repository by SBOM Tool and Language} 
    \label{fig:package_count_distribution}
\end{figure}

\textbf{Finding 1. \ghtool detects significantly more components per project than all other SBOM generation tool, except in Go projects.} Figure~\ref{fig:package_count_distribution} shows clear variation in the number of components detected across the four tools. 
In many ecosystems, including C\#, JavaScript, PHP, and Python, the \ghtool reports the highest average number of components. 
Trivy consistently reports the fewest components across most languages, while Syft and the \mstool generally fall in the middle.

\textbf{Finding 2. Go projects show consistency in the component counts across all SBOM tools.}
In our analysis using the Kruskal--Wallis H-test, we found no statistically significant differences in the average number of components of Go projects detected across all SBOM tools ($p = 0.186$). 
Likely due to Golang native support in listing software components, all four SBOM tools show a consistent average number of components per project. 
In contrast, projects from all other languagues exhibit statistically significant differences in detected component counts ($p < 0.05$).

\textbf{Finding 3. Pairwise comparisons reveal significant differences in component consistency across tools.}
In the nine programming languages with significant differences, we performed Dunn's post-hoc test with Bonferroni correction to examine pairwise differences between the \ghtool and all the other tools.

The pairwise analyses revealed the following patterns:

\begin{itemize}
    \item \textbf{\ghtool vs \mstool}: Both tools show no consistency on the number of components per project in all the remaining nine languages ($p < 0.05$), indicating significant differences in component detection between both tools.
    
    \item \textbf{\ghtool vs Syft}: Both tools showed comparable component count distribution in Java ($p = 0.466$) and Rust ($p = 0.601$). 
    All other languages showed significant differences in the distribution of components.
    
    \item \textbf{\ghtool vs Trivy}: Only Java projects showed consistency in the number of components between both tools ($p = 0.645$). 
    The remaining eight languages showed significant differences in the reported distributions of components.
    
\end{itemize}

Surprisingly, only SBOMs generated for Go projects and Java projects showed some level of consistency across tools. 
Across most of our dataset, \ghtool shows a statistically significant difference in the reported number of components if compared to the \mstool, Syft and Trivy.

\begin{figure}
    \centering
    \includegraphics[width=0.90\linewidth]{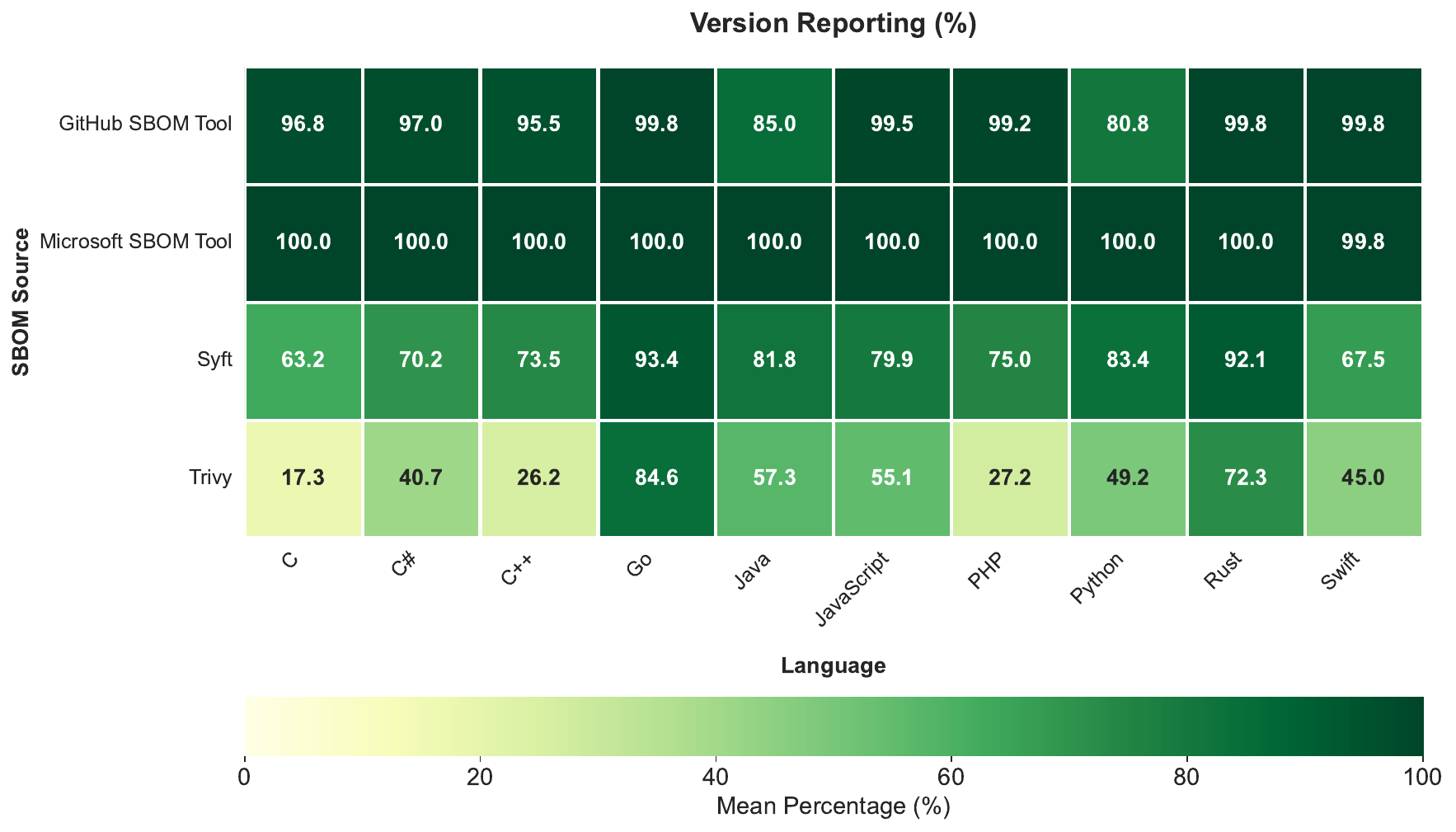}
    \caption{Average percentage of versioned components per project across languages and SBOM tools.} 
    
    \label{fig:version_heatmap}
\end{figure}

\textbf{Finding 4. \mstool consistently provides complete component versioning (100\%), while \ghtool reports 94\% of versioned components.}
We present in Figure~\ref{fig:version_heatmap} a heatmap of the average versioned component per project.
We notice that the \mstool provides version information for nearly every component across all languages. 
The \ghtool follows closely, reporting version data for almost all components except in Java (85\%) and Python projects (80\%). Syft performs well for Go and Rust projects, where approximately 90\% of components contain version information, but for projects in other languages, its coverage decreases to the 60--80\% range. 
Trivy shows the weakest performance overall, never reaching 90\% version reporting in any language and in some cases providing version data for only 15--30\% of components per project.

\begin{figure}
    \centering
    \includegraphics[width=1\linewidth]{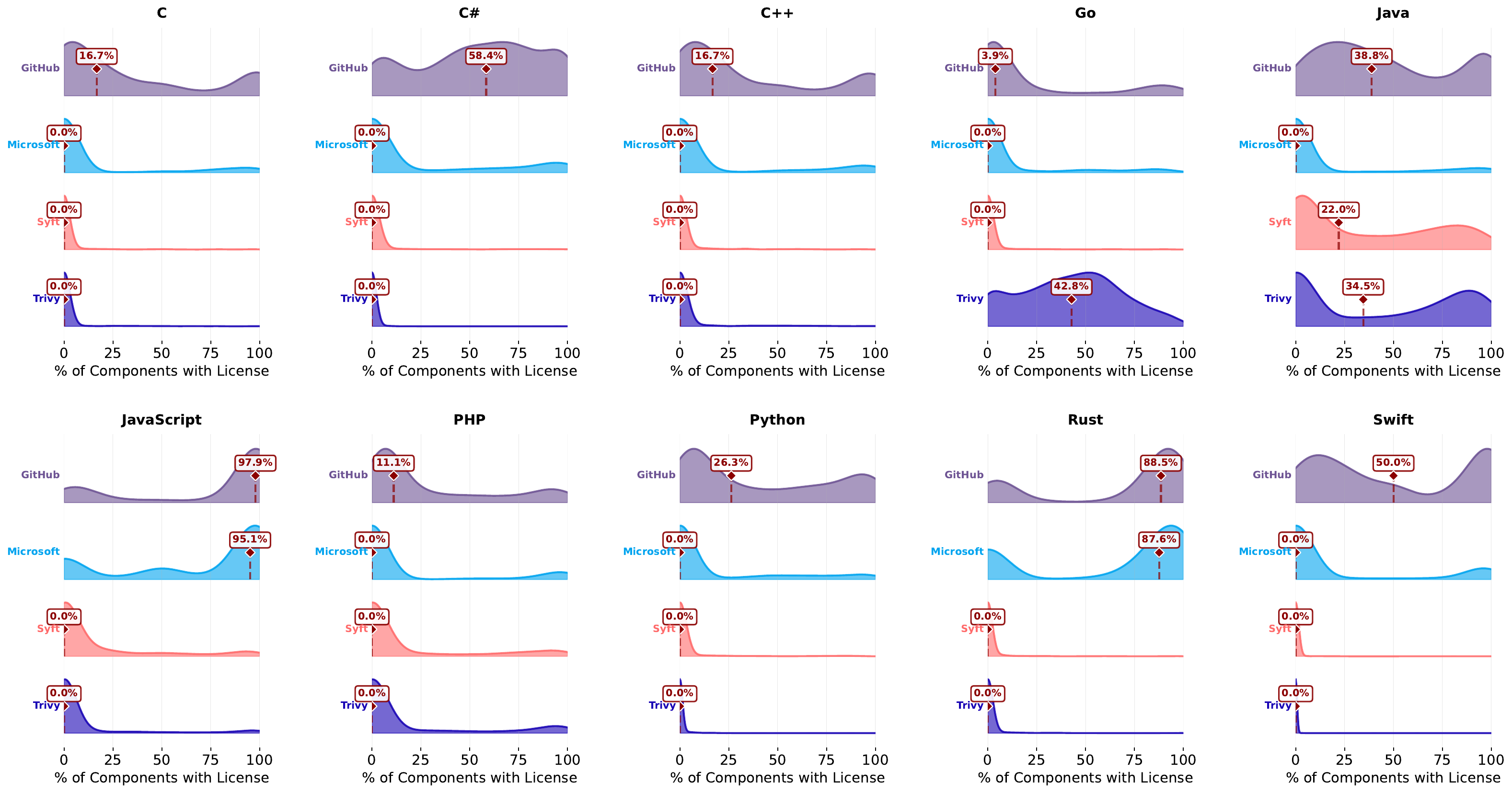}
    \caption{Distribution of License Availability Percentage Per Project by SBOM Tool and Language} \label{fig:license_percentage_distribution}
\end{figure}

\textbf{Finding 5. \ghtool provides more complete component license information than other tools for most programming languages.}
We show in Figure~\ref{fig:license_percentage_distribution} a distribution of the license availability percentage per project, across all tools and programming languages. 
We notice that \ghtool provides a more complete license information in nine out of ten languages, showin higher averages. However, even in these cases, coverage remains far from complete: only in Rust and JavaScript projects does the \ghtool achieve an average of more than 80\% of components' license information.
The \mstool delivers comparable performance to GitHub for JavaScript and Rust projects. For Java projects, Trivy and Syft perform similarly to the \ghtool. Go projects stand out as a notable exception: Trivy substantially outperforms all other tools by a wide margin.

\section{RQ3: \rqthree}
\label{sec:rq3}
Vulnerability detection is one of the most critical use cases of SBOMs. 
Modern software ecosystems contain large numbers of third-party libraries, and even a single outdated or vulnerable component can introduce significant security risks~\cite{Samaana_2025,jafari2022depsmells}. 
SBOMs are intended to support this process by providing the information necessary to match components against vulnerability databases and evaluate their security status. 
Hence, in this RQ, we try to assess how useful \ghtool is in detecting vulnerabilities in a Software product
and compare its performance with the three other widely used SBOM generation tools.

\subsection{Approach}

\noindent\textbf{PURL-Based Vulnerability Querying.}
To track software components, we rely on Package URLs (PURLs), which provide a canonical and ecosystem-agnostic mechanism for uniquely identifying software components.
We use OSV.dev, a distributed vulnerability database for open-source software, as our vulnerability data source~\cite{osvDev2025}. 
OSV.dev aggregates vulnerability information from multiple authoritative upstream sources and supports vulnerability retrieval through a public API. 
Vulnerabilities can be queried either by specifying a \textit{component} (or \textit{package})'s name and ecosystem or by providing a PURL. In our analysis, we exclusively use PURLs to ensure precise and consistent component matching.

We process all SBOMs generated by the four tools and attempt to extract PURLs for every component listed in each SBOM. 
For each SBOM, we examine the entry corresponding to every component and check for the presence of an \texttt{externalRefs} array, search for an item with \texttt{referenceType} set to \texttt{purl} and and extract the associated \texttt{referenceLocator} value, which contains the Package URL (PURL) of the component.
After collecting all available PURLs within a given SBOM, we submit them together to the batch query endpoint of the \texttt{osv.dev} API. The API returns all known vulnerabilities associated with each queried component, and we aggregate these results to obtain the complete set of vulnerabilities corresponding to the components listed in that SBOM.
\\

\noindent\textbf{Vulnerability Count Analysis.}
We first analyzed the total number of vulnerabilities associated with each SBOM and examined the distribution of vulnerability counts reported. 
Similar to the component count analysis (Section~\ref{sec:rq2}), we used the Kruskal–Wallis H-test and Dunn’s post-hoc test with Bonferroni correction to compare the distribution of vulnerability counts across SBOM tools.

\subsection{Results}

\textbf{Finding 1. PURL reporting is consistent in \ghtool and \mstool, while \syft and \trivy show inconsistency.} 
GitHub SBOMs provides a PURL for every component across all languages (100\% coverage). The \mstool performs similarly, reporting PURLs for nearly all listed components, with a negligible drop observed only for Swift (99.8\%).

Syft reports PURLs for a substantially smaller fraction of the listed components, and its performance varies notably by language. On average, Syft reports PURLs for approximately 60–90\% of the listed components for all the languages.
Trivy reports PURLs for the smallest fraction of listed components overall. For projects written in certain languages, including C (17.7\%), C++ (27.1\%), and PHP (27.3\%), fewer than one-third of the listed components include a PURL. These results indicate that, unlike GitHub and Microsoft, Syft and especially Trivy frequently omit PURLs for a large portion of the components they list.

\textbf{Finding 2. PURL availability is highest for Go and Rust projects.} Across all tools, Go and Rust projects show the highest PURL availability. Both GitHub and \mstool report PURLs for every component in Go and Rust projects. Syft and Trivy, despite their overall inconsistent performance, also perform relatively well for Go and Rust, achieving their highest PURL coverage in these projects. On average, Syft reports PURLs for 94.0\% of components in Go projects and 91.9\% in Rust projects. For Trivy, the values are 89.4\% in Go and 70.9\% in Rust projects.

\begin{figure}
    \centering
    \includegraphics[width=0.90\linewidth]{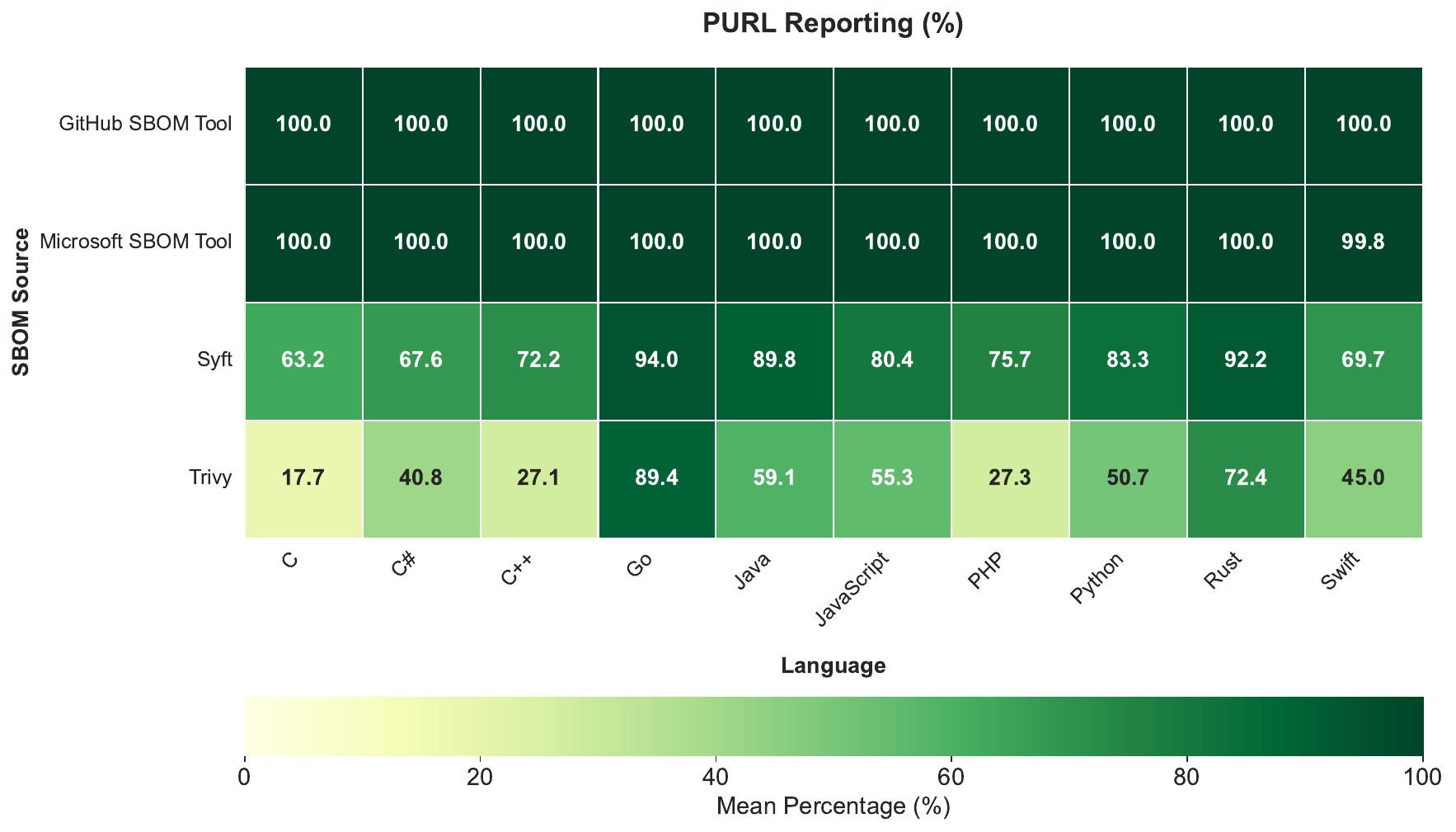}
    \caption{Average percentage of components with valid PURLs across languages and SBOM tools.} 
    \label{fig:purl_heatmap}
\end{figure}

\begin{figure}
    \centering
    \includegraphics[width=1\linewidth]{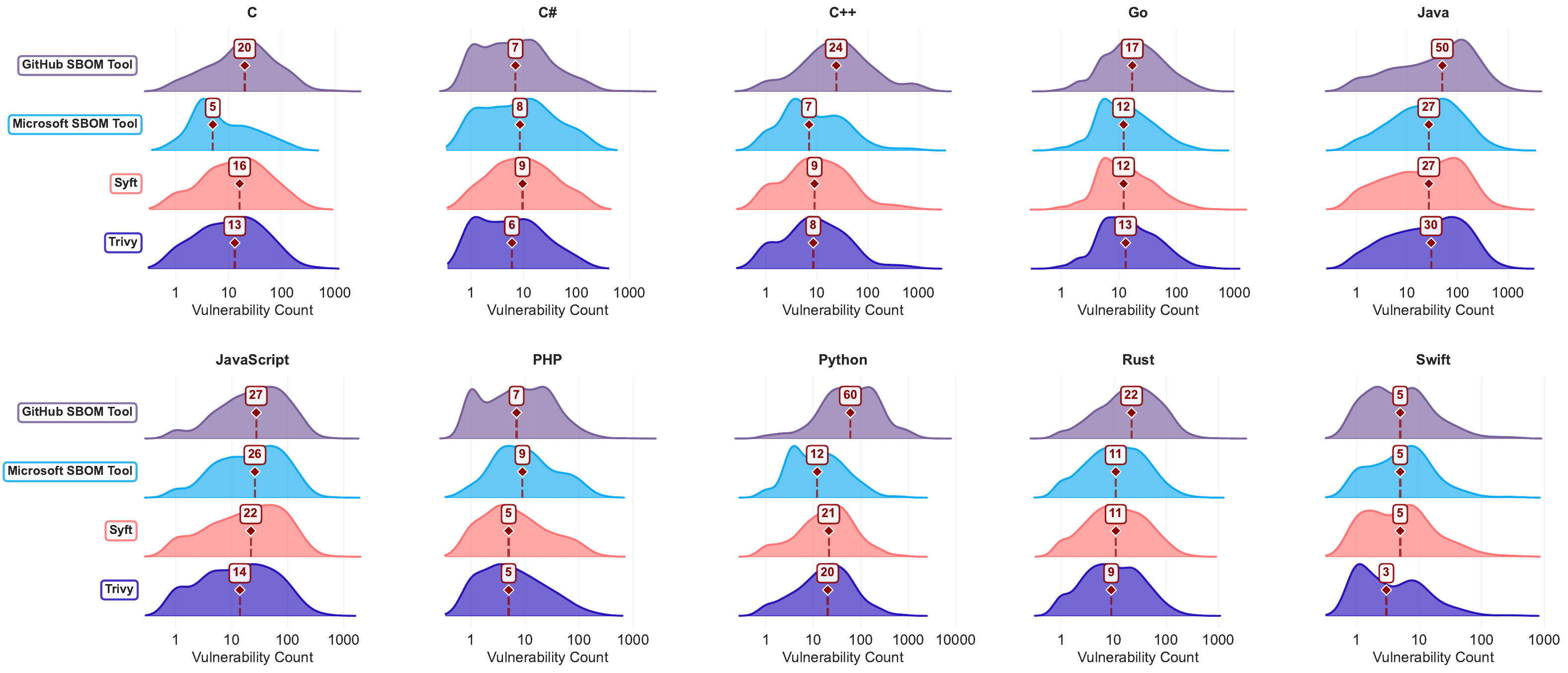}
    \caption{Vulnerability Count Distribution Per Repository by SBOM Tool and Language.} 
    \label{fig:vulnerability_count_distribution}
\end{figure}

\textbf{Finding 3. The number of vulnerabilities detected varies significantly by tool.} 
As shown in Figure~\ref{fig:vulnerability_count_distribution}, the \ghtool consistently detects a higher number of vulnerabilities than the other tools across almost all programming languages. 
This outcome was expected, as the \ghtool generally reports more components and provides a PURL for each listed component, enabling complete vulnerability lookups via OSV.dev.

Among the remaining tools, Syft and the \mstool report broadly similar vulnerability counts, though their relative differences vary by language. Syft reports more vulnerabilities in Python, C, and C++ projects, while the \mstool reports more in JavaScript and PHP projects. Trivy consistently reports the lowest number of vulnerabilities across nearly all languages. This can be attributed to its combination of listing fewer components and frequently omitting PURLs for the components it does report, which limits the effectiveness of OSV. dev-based queries.

Applying the Kruskal–Wallis H-test to the vulnerability counts for each programming language, we found that all ten languages showed statistically significant differences ($p < 0.05$) in vulnerability detection across the four SBOM tools.
Dunn’s post-hoc test with Bonferroni correction to examine pairwise differences in vulnerability detection between the \ghtool and each competing tool revealed the following patterns:

\begin{itemize}
    \item \textbf{\ghtool vs \mstool}: The number of vulnerabilities reported between both tools was consistent in C\# ($p = 0.224$), JavaScript ($p = 0.112$), and Swift projects ($p = 0.421$). 
    All other languages showed statistically significant differences. 
    
    \item \textbf{\ghtool vs Syft}: Both tools showed a consistent number of vulnerabilities only in C ($p = 0.127$), PHP ($p = 0.353$) and Swift projects ($p = 0.944$). 
    Projects from all other languages showed a distinct number of vulnerabilities.
    
    \item \textbf{\ghtool vs Trivy}: 
    Both tools showed a consistent number of vulnerabilities in projects written in C\# ($p = 0.224$) and PHP ($p = 0.052$).
    Tools significantly diverge in all other remaining languages.
    
\end{itemize}

While no single programming language showed consistent vulnerability counts across all tools, \ghtool is sporadically consistent with other tools in specific languages, usually related to C\#, PHP and Swift. 

\begin{figure}
    \centering
    \includegraphics[width=\linewidth]{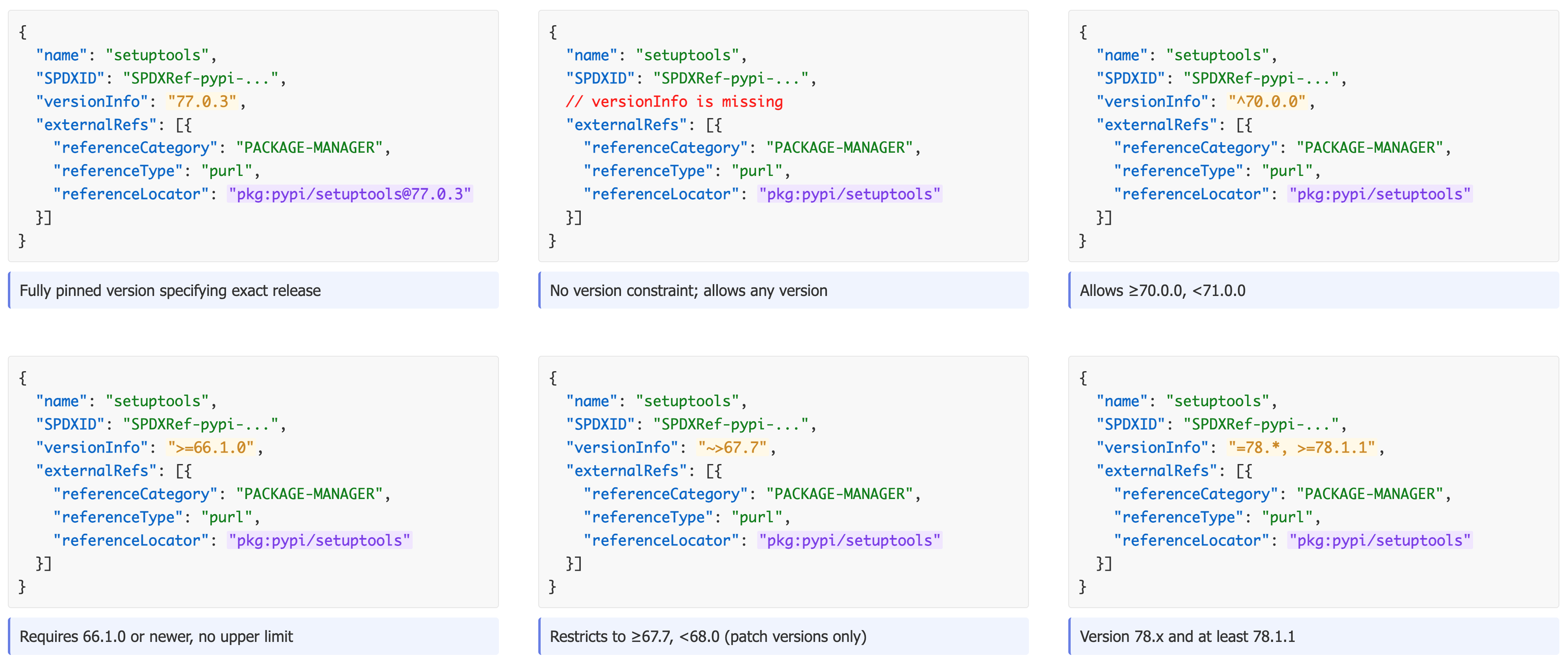}
    \caption{Effect of presence of version information on PURL.} 
    \label{fig:version_vs_purl}
\end{figure}

\begin{figure}
    \centering
    \includegraphics[
        width=0.95\linewidth,
        trim=1cm 8.5cm 5cm 2cm,
        clip
    ]{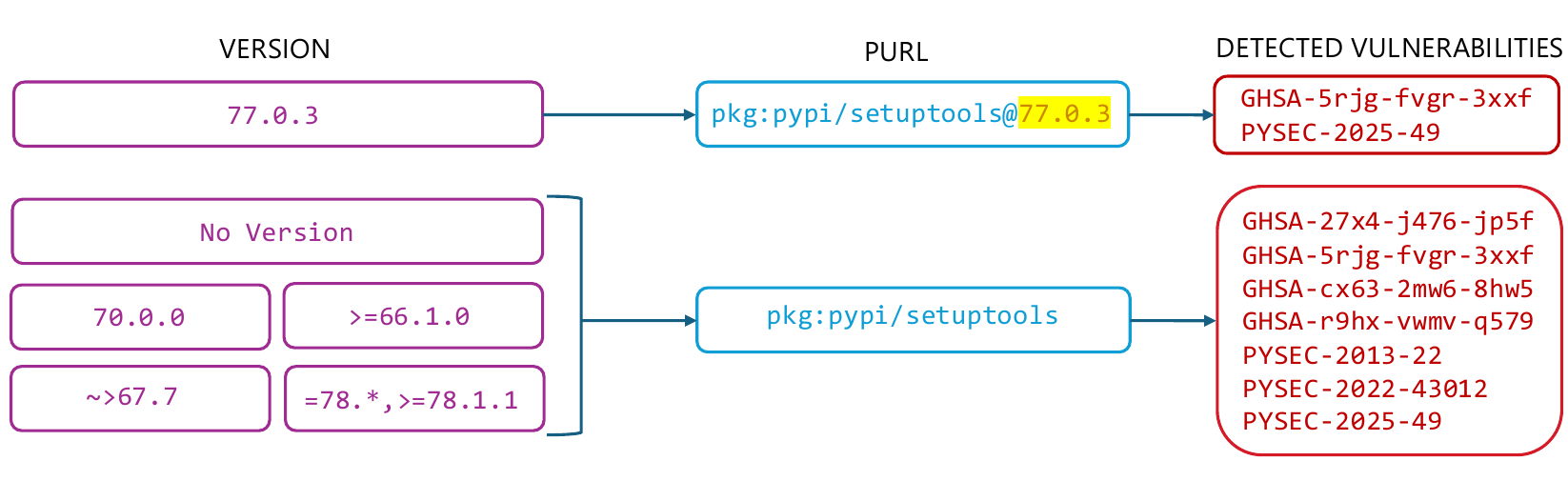}
    \caption{How version information's availability in PURL affects detection of vulnerabilities.} 
    \label{fig:vesion_to_purl_to_vulnerability}
\end{figure}

\noindent
\textbf{A caveat about the number of vulnerabilities.}
The precision of vulnerability detection depends on the availability of version information.
Our analysis indicates that while OSV queries based on PURLs always return vulnerability information for a given dependency, the precision of these results critically depends on whether the PURL specifies an exact version.
If the \ghtool does not include an exact version for a component, the resulting PURL is versionless. Figure~\ref{fig:version_vs_purl} illustrates six different ways the component \texttt{setuptools} appears across the SBOMs. It can be observed that only when an exact version—\texttt{77.0.3} in this case—is provided does the PURL reflect that version; in all other cases, where no version or a non-exact version is provided, the PURL is versionless.

Submitting a versionless PURL to the \texttt{osv.dev} API returns all vulnerabilities associated with that component, regardless of version. For example, as shown in Figure~\ref{fig:vesion_to_purl_to_vulnerability}, querying with the PURL \texttt{pkg:pypi/setuptools@77.0.3} returns two vulnerabilities, enabling the software consumer to identify exactly which vulnerabilities affect their software. In contrast, querying with the versionless PURL \texttt{pkg:pypi/setuptools} returns six vulnerabilities, leaving the consumer uncertain about which vulnerabilities are relevant to their software.

\section{Discussion}
\label{sec:discussion}

In this section, we discuss four different aspects of our results.
We first present an accuracy evaluation of the four SBOM tools against a manually constructed ground truth for Python, Java, and JavaScript (Section~\ref{sub:groundtruth}). We then compare the component agreement level between the \ghtool and the \mstool (Section~\ref{sub:gh-vs-ms}), discuss the lack of component versioning in Python projects (Section~\ref{sub:python-indepth}), and examine the lack of license information across ecosystems (Section~\ref{sub:licensing-discussion}).

\subsection{Accuracy Evaluation Against Ground Truth}
\label{sub:groundtruth}

The analyses described in RQ2 measure the \emph{presence} of metadata fields in generated SBOMs, but do not assess whether the reported values are \emph{correct}.
A tool may report an incorrect version or omit the license, undermining further compliance analysis, or fail to report components that the repository actually uses, introducing false negatives that leave real dependencies invisible to downstream consumers. 

To address this, we construct a ground truth dataset for Python, Java, and JavaScript, three languages whose package ecosystems provide programmatic access to authoritative dependency, version, and license metadata, enabling reliable automated ground truth. 
Given the time-consuming task of building projects to resolve their actual dependency tree, we randomly sampled 200 repositories per language from our full dataset,
yielding 600 repositories in total.

For constructing the ground truth, we resolve each repository's dependencies programmatically from each ecosystem's native tooling.

\begin{itemize}
    \item For Python, following the methodology of Yu et al.~\cite{Yu2024CorrectnessSBOM}, we install each repository's dependencies into a fresh virtual environment using \texttt{pip install} and then capture the full resolved environment via \texttt{pip freeze}. This produces the complete set of transitive dependencies, each annotated with its exact pinned version.

    \item For Java, we run \texttt{mvn dependency:tree} against each repository's \texttt{pom.xml}, which triggers Maven's dependency resolver and returns the fully resolved transitive dependency tree, including the concrete version selected for each artifact.
    
    \item For JavaScript, we parse \texttt{package-lock.json} or \texttt{yarn.lock}, whichever is present in the repository, to extract the full set of installed packages together with their locked versions. During repository sampling, only repositories containing at least one of these two lockfiles were selected; repositories lacking both lockfiles were excluded and replaced by the next sampled candidate. 

\end{itemize}

For license ground truth, we query each ecosystem's authoritative registry: the PyPI~\cite{pypi} API for Python; the npm~\cite{npm} registry API for JavaScript; and the Libraries.io~\cite{librariesio} API for Java, since Maven Central does not expose a structured REST API for license data and its POM \texttt{<licenses>} field is free-form text that is unsuitable for automated normalisation.

We evaluate tools at the \textbf{component detection} level (Precision, Recall, F1 via normalized name matching) and the \textbf{version accuracy} and \textbf{license accuracy} levels (fraction of True Positives where the reported value exactly matches the ground truth, macro-averaged across repositories). For version accuracy, we require an exact match between the version reported in the SBOM and the version present in the ground truth. A tool that reports a version range or a resolved value that differs from the ground truth is considered incorrect.

\begin{table*}[t]
\centering
\caption{Component detection, version reporting accuracy, and license reporting accuracy
         evaluated against ground truth, macro-averaged across repositories.
         Best value per metric per language is shown in \textbf{bold}.}
\label{tab:ground_truth_accuracy}

\resizebox{\textwidth}{!}{%
\begin{tabular}{ll rrrrr}
\toprule
\textbf{Language} & \textbf{Tool} &
\textbf{\makecell{Component\\Detection\\Precision}} &
\textbf{\makecell{Component\\Detection\\Recall}} &
\textbf{\makecell{Component\\Detection\\F1}} &
\textbf{\makecell{Version\\Reporting\\Accuracy}} &
\textbf{\makecell{License\\Reporting\\Accuracy}} \\
\midrule
\multirow{4}{*}{Python}
  & \ghtool    &          33.7\% &          57.3\% &          29.0\% &          21.6\% & 19.7\% \\
  & \mstool    & \textbf{44.4\%} & \textbf{66.4\%} & \textbf{47.9\%} & \textbf{69.8\%} & \textbf{27.6\%} \\
  & Trivy      &          20.3\% &          32.3\% &          19.8\% &          27.4\% &  1.7\% \\
  & Syft       &          13.2\% &          34.1\% &          13.2\% &          31.0\% &  3.9\% \\
\midrule
\multirow{4}{*}{Java}
  & \ghtool    &          46.8\% &          38.2\% &          35.9\% &          53.7\% & 28.5\% \\
  & \mstool    & \textbf{79.3\%} & \textbf{98.4\%} & \textbf{84.6\%} & \textbf{95.6\%} &  8.7\% \\
  & Trivy      &          55.7\% &          54.9\% &          50.5\% &          81.9\% & \textbf{60.2\%} \\
  & Syft       &          55.7\% &          36.9\% &          38.7\% &          73.5\% &  1.3\% \\
\midrule
\multirow{4}{*}{JavaScript}
  & \ghtool    & \textbf{88.2\%} & \textbf{86.7\%} & \textbf{85.3\%} & \textbf{90.4\%} & \textbf{94.7\%} \\
  & \mstool    &          87.9\% &          80.3\% &          81.1\% &          88.8\% & 78.5\% \\
  & Trivy      &          65.1\% &          26.5\% &          30.1\% &          66.9\% &  1.3\% \\
  & Syft       &          63.8\% &          40.0\% &          41.2\% &          69.2\% & 25.6\% \\
\bottomrule
\end{tabular}%
}

\end{table*}

Table~\ref{tab:ground_truth_accuracy} presents the component detection, version accuracy, and license accuracy results evaluated against our ground truth dataset across Python, Java, and JavaScript. While RQ2 consistently positioned the \ghtool as the leading tool in terms of component count and metadata presence, the accuracy evaluation reveals a more nuanced picture. For component detection, the \ghtool's higher component count in Python and Java does not translate into higher accuracy against the ground truth; the \mstool, despite reporting fewer components, achieves better precision, recall, and F1 in both ecosystems. For version accuracy, the \mstool's near-complete version presence observed in RQ2 does carry over into correctness: it leads on version accuracy in two of the three ecosystems, confirming that it not only populates version fields consistently but also reports the right values. For license accuracy, three different tools lead in three ecosystems.

In \textbf{Python}, the \mstool leads on all component detection metrics, achieving a Precision of 44.4\%, Recall of 66.4\%, and F1 of 47.9\%, alongside the highest version accuracy of 69.8\% and license accuracy of 27.6\%. This is somewhat surprising given that RQ2's Finding~1 (Section~\ref{subsec:rq2_findings}) identified the \ghtool as detecting more components than the \mstool in Python in terms of raw component counts. However, a higher component count does not imply higher accuracy: the \ghtool surfaces more components overall, but fewer of them match the ground truth, resulting in lower component detection performance than the \mstool. Version accuracy is low across all tools except the \mstool, ranging from 21.6\% (\ghtool) to 31.0\% (Syft). The \mstool's relatively high version accuracy of 69.8\% suggests that it resolves undeclared or range-based versions more reliably than other tools while the other tools may provide range-based versions or wrong resolved value.
License accuracy is uniformly poor across all tools, with no tool exceeding 27.6\%.

In \textbf{Java}, the \mstool dominates component detection with a Precision of 79.3\%, Recall of 98.4\%, and F1 of 84.6\%, and achieves the highest version accuracy at 95.6\%. RQ2's Finding~4 showed that \mstool provides version information for every detected component across nearly all ecosystems, and the 95.6\% version accuracy confirms that these reported versions are also almost always correct for Java projects. The \ghtool performs considerably weaker in component detection (F1 of 35.9\%) and version accuracy (53.7\%), consistent with our earlier observation in 
RQ1's Finding~2 (Section~\ref{sec:rq1}) that GitHub SBOMs omit transitive dependencies of unversioned parent components. Notably, the \mstool's dominant component detection performance does not extend to license accuracy, where it achieves only 8.7\% compared to \ghtool's 28.5\% or Trivy's 60.2\%.

In \textbf{JavaScript}, the \ghtool leads across all metrics, achieving a Precision of 88.2\%, Recall of 86.7\%, F1 of 85.3\% for component detection, version accuracy of 90.4\%, and license accuracy of 94.7\%.
The strong version accuracy of the \ghtool in JavaScript suggests that it benefits from the availability of lockfiles, which provide exact pinned versions and remove the need to resolve range-based declarations. The \mstool is competitive across component detection and version accuracy (81.1\% and 88.8\% respectively) but falls more behind on license accuracy (78.5\%). 
The license accuracy results for both tools are consistent with RQ2's Finding~5 (Section~\ref{subsec:rq2_findings}), which identified JavaScript as one of only two languages where the \ghtool achieves above 80\% license presence and \mstool also reports similar performance for JavaScript projects. Our ground truth analysis confirms that the reported licenses are not only present but also largely correct. Trivy and Syft show sharply lower performance in component detection, particularly on recall (26.5\% and 40.0\%), confirming that they detect a selective subset of the full dependency tree in this ecosystem.

\subsection{GitHub SBOM vs MS SBOM Generation tool}
\label{sub:gh-vs-ms}

Among the SBOM generation tools evaluated, both the \mstool and the \ghtool demonstrated reliable performance in providing component version and license information (RQ2) and PURL information (RQ3). 
However, high completeness scores do not preclude the possibility that the two tools are detecting substantially different sets of dependencies.

To examine whether the tools also agree on the components they report, we perform a component-level agreement analysis. For the component-level comparison, each component's PURL is normalized by removing the version from the PURL, causing multiple versions of the same component to collapse into a single entry; we refer to these as \textit{unique components}. For the versioned comparison, each entry is keyed by the pair (versionless PURL, \texttt{versionInfo}), such that the same component appearing with different versions is treated as distinct entries; we refer to these as \textit{unique versioned components}. Packages lacking version information are excluded from the versioned comparison.

Table~\ref{tab:agreement} shows the percentage of components that are uniquely reported in GitHub SBOMs (GH), reported in both GitHub and Microsoft SBOMs (GH $\cap$ MS), and uniquely reported to Microsoft (MS). 
In this analysis, we first summed all three component categories per project and computed the percentage of the aggregated sum across all unique components found in both tools' SBOMs. 

Overall, in most languages \ghtool and \mstool tend to agree on the majority of unique components.
The highest agreement is observed in Go (76.8\%), JavaScript (75.0\%), and Rust (72.8\%) projects, suggesting strong consistency. At the same time, GitHub tends to identify more components in 8 out of 10 languages, with Java and Rust projects being the only exceptions. 
In contrast, PHP exhibits the lowest overlap (43.0\%), where more than half of the components (50.5\%) are reported exclusively by GitHub, indicating substantial divergence between the tools.

When accounting for versioned components, i.e., whether both \ghtool and \mstool report the same component and their versions, the agreement decreases. For example, in Go projects, agreement drops from 76.8\% (components) to 32.6\% (versioned components), with both tools reporting roughly one-third of versioned components exclusively. A similar decline is observed in Java and C++ projects, suggesting that version resolution and normalization strategies differ significantly. Java, in particular, shows a notable asymmetry, with Microsoft reporting 41.1\% of versioned components exclusively compared to 22.8\% for GitHub.
These results indicate that while the tools often agree on the presence of dependencies at the component level, discrepancies become more pronounced when version information is incorporated. 

\begin{table}[]
    \centering
    \caption{Agreement level between \ghtool (GH) and the \mstool (MS). We highlight in light grey values above 25\% and in dark grey values above 50\%.}
    \label{tab:agreement}
    
\resizebox{\linewidth}{!}{%
\begin{tabular}{l|*{3}{w{c}{1.5cm}}|*{3}{w{c}{1.5cm}}}
\toprule
& \multicolumn{6}{c}{\textbf{Agreement Level}} \\
\textbf{Languages} & \multicolumn{3}{c}{\textbf{Unique Components}}
                    & \multicolumn{3}{c}{\textbf{Unique Versioned Components}} \\
                    & GH & GH$\cap$MS & MS
                    & GH & GH$\cap$MS & MS \\
\midrule
C          & \cellcolor{black!10} 28.3\% & \cellcolor{black!25} 52.5\% & 19.1\% & \cellcolor{black!10} 33.4\% & \cellcolor{black!10} 43.9\% & 22.8\% \\
C\#        & \cellcolor{black!10} 38.2\% & \cellcolor{black!25} 52.7\% &  9.1\% & \cellcolor{black!10} 42.4\% & \cellcolor{black!10} 43.3\% & 14.3\% \\
C++        & \cellcolor{black!10} 27.5\% & \cellcolor{black!25} 53.7\% & 18.8\% & \cellcolor{black!10} 33.4\% & \cellcolor{black!10} 41.7\% & 24.9\% \\
Go         & 13.0\% & \cellcolor{black!25} 76.8\% & 10.3\% & \cellcolor{black!10} 34.0\% & \cellcolor{black!10} 32.6\% & \cellcolor{black!10} 33.4\% \\
Java       & 17.6\% & \cellcolor{black!10} 49.2\% & \cellcolor{black!10} 33.1\% & 22.8\% & \cellcolor{black!10} 36.1\% & \cellcolor{black!10} 41.1\% \\
JavaScript & 15.1\% & \cellcolor{black!25} 75.0\% &  9.9\% & 22.4\% & \cellcolor{black!25} 62.4\% & 15.2\% \\
PHP        & \cellcolor{black!25} 50.5\% & \cellcolor{black!10} 43.0\% &  6.5\% & \cellcolor{black!25} 51.8\% & \cellcolor{black!10} 36.4\% & 11.8\% \\
Python     & 24.2\% & \cellcolor{black!25} 57.9\% & 17.9\% & \cellcolor{black!10} 29.6\% & \cellcolor{black!10} 45.9\% & 24.5\% \\
Rust       & 13.2\% & \cellcolor{black!25} 72.8\% & 14.0\% & \cellcolor{black!10} 29.1\% & \cellcolor{black!10} 44.6\% & \cellcolor{black!10} 26.3\% \\
Swift      & 20.5\% & \cellcolor{black!25} 66.1\% & 13.3\% & 23.2\% & \cellcolor{black!25} 60.0\% & 16.8\% \\
\bottomrule
\end{tabular}%
}

\end{table}

\subsection{Potential causes for lack of versioning of Python dependency components. }
\label{sub:python-indepth}

From the results of RQ2, we observe that \ghtool provides low support for component versioning in Python projects, with an average of 80\% of components per project being versioned. 

To understand the potential causes, we automatically analyzed the \\ \texttt{requirements.txt} files of all Python projects in our dataset. Python allows developers to import and use a library without specifying it in any metadata file, as long as it is installed in the runtime environment. Furthermore, Python projects offer multiple mechanisms for specifying dependencies, including \texttt{requirements.txt}, \texttt{setup.py}, \texttt{pyproject.toml}, and \texttt{Pipfile}, among others --- though our analysis focused on \texttt{requirements.txt} files, which are the most widely used convention for specifying installation requirements. Among the 9,575 PyPI library instances missing version information in \ghtool-generated SBOMs, we identified distinct causes.

\begin{itemize}
    \item \textbf{Lack of version specifiers.} In 5,961 cases (62.26\%), the libraries were present in requirements.txt files but declared without any version specifier, which directly propagates to the SBOM as an unversioned entry. 
    
    \item \textbf{Absent libraries in the manifest file.} In the remaining 3,614 cases (37.74\%), the libraries were absent from requirements.txt entirely, attribu-table to one of three scenarios: the dependency may be declared in an alternative metadata file not covered by our analysis; it may be a transitive dependency resolved by GitHub from the full dependency tree; or it may not be declared anywhere at all, with \ghtool surfacing it in the SBOM through static analysis of import statements in the source code.
\end{itemize}

These findings suggest that the version-adherence problem in GitHub\\
SBOMs is primarily due to loose dependency-declaration practices in the Python ecosystem rather than a limitation of the SBOM generation tool itself. 

\subsection{Why is licensing information so seldom reported in SBOMs?}
\label{sub:licensing-discussion}

Our results show that component license information is seldom reported in SBOMs, across all tools. 
This may indicate that poor license coverage is not just a tool limitation, but likely reflects a limitation on how license metadata is resolved.
We break the analysis of the license into two orthogonal factors: 1) dependency versioning practices and 2) the related package manager ecosystem.
To this aim, we classify every dependency exported across all SBOMs analyzed based on their reported version into three categories: 

\begin{itemize}
    \item \textbf{Exact version.} We classify a dependency as having an exact version if the version is expressed in the SBOM as a unique version, without ranges or semantic versioning, a.k.a., dependency pinning. 

    \item \textbf{Non-exact version}. We classify a dependency's version as \textit{Non-exact} if it is expressed as a range, a wildcard, or using semantic versioning (semver) operators.

    \item \textbf{No version}. Dependencies where no version is specified in the SBOM. 
\end{itemize}

\begin{table*}
\caption{License Availability Across Package Managers Based on Version Specification.}
\label{tab:license_vs_version_across_pms}
\centering
\renewcommand{\arraystretch}{1.2}
\small

\resizebox{\linewidth}{!}{%
\begin{tabular}{l r r r r r r r}
    \toprule
    Package Manager & \multicolumn{1}{c}{Total} & \multicolumn{2}{c}{Exact Version} & \multicolumn{2}{c}{Non-Exact Version} & \multicolumn{2}{c}{No Version} \\
    & Dependencies & Dependencies & Licensed \% & Dependencies & Licensed \% & Dependencies & Licensed \% \\
    \midrule
    pypi          &    82,191  &   54,647  & 86.20\%   &   12,679  & \cellcolor{black!10} 0.00\%   &  14,865  & \cellcolor{black!10} 0.00\%   \\
    cargo         &   353,825  &  319,231  & 98.71\%   &   34,594  & \cellcolor{black!10} 0.00\%   &       0  & \cellcolor{black!10} 0.00\%   \\
    githubactions &    58,953  &   58,906  & \cellcolor{black!10} 0.00\%   &       47  & \cellcolor{black!10} 0.00\%   &       0  & \cellcolor{black!10} 0.00\%   \\
    maven         &   114,587  &   74,150  & 48.21\%   &      226  & \cellcolor{black!10} 0.00\%   &  40,211  & \cellcolor{black!10} 0.00\%   \\
    nuget         &    43,206  &   31,809  & 66.56\%   &    9,556  & \cellcolor{black!10} 0.00\%   &   1,841  & \cellcolor{black!10} 0.00\%   \\
    npm           & 2,708,803  & 2,647,719 & 99.18\%   &   61,070  & \cellcolor{black!10} 0.00\%   &      14  & \cellcolor{black!10} 0.00\%   \\
    golang        &   128,770  &  128,770  & \cellcolor{black!10} 0.00\%   &        0  & \cellcolor{black!10} 0.00\%   &       0  & \cellcolor{black!10} 0.00\%   \\
    gem           &    25,048  &   24,258  & 97.34\%   &      790  & \cellcolor{black!10} 0.00\%   &       0  & \cellcolor{black!10} 0.00\%   \\
    composer      &    41,970  &   30,041  & 41.76\%   &   11,501  & \cellcolor{black!10} 0.00\%   &     428  & \cellcolor{black!10} 0.00\%   \\
    pub           &     3,333  &    2,872  & \cellcolor{black!10} 0.00\%   &      461  & \cellcolor{black!10} 0.00\%   &       0  & \cellcolor{black!10} 0.00\%   \\
    swift         &     3,854  &    3,854  & \cellcolor{black!10} 0.00\%   &        0  & \cellcolor{black!10} 0.00\%   &       0  & \cellcolor{black!10} 0.00\%   \\
    github        &       208  &      208  & \cellcolor{black!10} 0.00\%   &        0  & \cellcolor{black!10} 0.00\%   &       0  & \cellcolor{black!10} 0.00\%   \\
    deb           &       589  &      589  & \cellcolor{black!10} 0.00\%   &        0  & \cellcolor{black!10} 0.00\%   &       0  & \cellcolor{black!10} 0.00\%   \\
    \bottomrule
\end{tabular}%
}

\end{table*}

Table~\ref{tab:license_vs_version_across_pms} presents license availability across package
managers, broken down by component version strategy~\cite{jafari2022depsmells}. 
We observe a clear pattern across all package managers: dependencies reported with an exact version are far more likely to include license information, whereas those with missing or non-exact versions consistently lack versions. 
We also observe that exact versioning alone does not guarantee license availability. For dependencies managed by the package managers \texttt{swift}, \texttt{pub}, \texttt{deb}, \texttt{golang}, \texttt{github}, and \texttt{githubactions}, the tool fails to retrieve license information for all dependencies, regardless of the versioning strategy. 
Based on this, we examined all unlicensed dependencies per language and attributed each to one of the following probable causes:

\begin{itemize}
    \item \textbf{Explanation-1 (E1):} Dependency is found in the SBOM having a non-exact version or having no version, e.g., dependencies declared using semantic versioning.  

    \item \textbf{Explanation-2 (E2):} Dependency is listed in the SBOM with a \texttt{pkg:\allowbreak githubactions} type, indicating it is a GitHub Action referenced within the project's CI/CD pipeline.
    \item \textbf{Explanation-3 (E3):} Dependency appears in the SBOM with a \texttt{pkg:\allowbreak golang} type, indicating it is a Golang component.
    \item \textbf{Explanation-4 (E4):} Dependency is listed in the SBOM with a \texttt{pkg:pub} type, indicating it is a Pub component.
    \item \textbf{Explanation-5 (E5):} Dependency is found in the SBOM with a \texttt{pkg:\allowbreak swift} type, indicating it is a Swift component.
    \item \textbf{Explanation-6 (E6):} Dependency appears in the SBOM with a \texttt{pkg:\allowbreak github} type, indicating it is a component sourced directly from a GitHub repository rather than a formal registry.
    \item \textbf{Explanation-7 (E7):} Dependency is listed in the SBOM with a \texttt{pkg:deb} type, indicating it is a Debian system-level component.
\end{itemize}

\begin{table}
    \caption{Explanations for lack of license information}
    \label{tab:lack-license}
    \centering
    \resizebox{\linewidth}{!}{%
\begin{tabular}{l|r r r r r r r r r}
\toprule
\textbf{Language} 
    & \makecell{\textbf{\#}\\[-2pt]\textbf{Unlicensed}} 
    & \makecell{\textbf{Non-Exact/}\\[-2pt]\textbf{No Version (E1)}} 
    & \makecell{\textbf{GH Actions}\\[-2pt]\textbf{(E2)}} 
    & \makecell{\textbf{Golang}\\[-2pt]\textbf{(E3)}} 
    & \makecell{\textbf{Pub}\\[-2pt]\textbf{(E4)}} 
    & \makecell{\textbf{Swift}\\[-2pt]\textbf{(E5)}} 
    & \makecell{\textbf{GitHub}\\[-2pt]\textbf{(E6)}} 
    & \makecell{\textbf{Deb}\\[-2pt]\textbf{(E7)}} 
    & \makecell{\textbf{Unac-}\\[-2pt]\textbf{counted}} \\
\midrule
C      & 11360  & 3101 (27.3\%)  & 4779 (42.1\%)  & 1869 (16.5\%)   & 637 (5.6\%)  & 15 (0.1\%)    & 0 (0.0\%)   & 0 (0.0\%)   & 959 (8.4\%)    \\
C\#    & 30050  & 13953 (46.4\%) & 4660 (15.5\%)  & 182 (0.6\%)     & 92 (0.3\%)   & 13 (0.0\%)    & 16 (0.1\%)  & 0 (0.0\%)   & 11134 (37.1\%) \\
C++    & 19344  & 7891 (40.8\%)  & 6584 (34.0\%)  & 1728 (8.9\%)    & 412 (2.1\%)  & 72 (0.4\%)    & 191 (1.0\%) & 220 (1.1\%) & 2246 (11.6\%)  \\
Go     & 135288 & 4135 (3.1\%)   & 9713 (7.2\%)   & 119099 (88.0\%) & 140 (0.1\%)  & 8 (0.0\%)     & 0 (0.0\%)   & 0 (0.0\%)   & 2193 (1.6\%)   \\
Java   & 96736  & 47195 (48.8\%) & 5805 (6.0\%)   & 1868 (1.9\%)    & 1022 (1.1\%) & 26 (0.0\%)    & 1 (0.0\%)   & 0 (0.0\%)   & 40819 (42.2\%) \\
JavaScript  & 44342  & 28766 (64.9\%) & 4219 (9.5\%)   & 1573 (3.5\%)    & 0 (0.0\%)    & 16 (0.0\%)    & 0 (0.0\%)   & 0 (0.0\%)   & 9768 (22.0\%)  \\
PHP    & 38607  & 14569 (37.7\%) & 4582 (11.9\%)  & 485 (1.3\%)     & 0 (0.0\%)    & 0 (0.0\%)     & 0 (0.0\%)   & 0 (0.0\%)   & 18971 (49.1\%) \\
Python & 40260  & 23634 (58.7\%) & 7423 (18.4\%)  & 988 (2.5\%)     & 337 (0.8\%)  & 2 (0.0\%)     & 0 (0.0\%)   & 369 (0.9\%) & 7507 (18.6\%)  \\
Rust   & 60916  & 44691 (73.4\%) & 8909 (14.6\%)  & 899 (1.5\%)     & 18 (0.0\%)   & 32 (0.1\%)    & 0 (0.0\%)   & 0 (0.0\%)   & 6367 (10.5\%)  \\
Swift  & 7025   & 348 (5.0\%)    & 2232 (31.8\%)  & 79 (1.1\%)      & 214 (3.0\%)  & 3670 (52.2\%) & 0 (0.0\%)   & 0 (0.0\%)   & 482 (6.9\%)    \\
\bottomrule
\end{tabular}}
    
\end{table}

Table~\ref{tab:lack-license} presents the result of our analysis across all ten languages. 
Whenever possible, each dependency was categorized in one of the explanations. 
Dependencies that could not be attributed to any
of the identified explanations are reported as \textit{Unaccounted}.

The most pervasive cause of missing license information is a non-exact or absent version
specification (E1). In 8 out of 10 languages, E1 is the single largest contributor to
unlicensed dependencies. 
Among dependencies that do have an exact version, missing licenses are explained by
membership in package managers for which the \ghtool does not retrieve license
information (E2--E7). 
The degree to which these causes account for the remaining missing
licenses varies substantially across languages. For Go, Golang dependencies (E3) alone account
for 88.0\% of all unlicensed dependencies, leaving only 1.6\% unaccounted for. Swift is
similarly well-explained, with Swift dependencies (E5) and GitHub Actions (E2) together
covering 84.0\% of missing licenses. In contrast, Java and PHP prove harder to explain:
even after accounting for all identified causes, 42.2\% and 49.1\% of their unlicensed
dependencies, respectively, remain unaccounted for --- meaning these dependencies carry exact
version information and belong to supported package managers, yet the \ghtool
still fails to retrieve their license metadata. This points to gaps in the upstream
registries the tool relies upon, and represents an important avenue for future
improvement.

\section{Related Works}
\label{sec:related_works}

In this section, we discuss the related works.

\subsection{Empirical Studies on SBOM.}
Empirical research on SBOM adoption has provided useful insights into the practices and challenges observed across industry and open-source ecosystems. Xia et al. conducted an empirical study based on interviews with practitioners and a survey of professionals to examine current SBOM practices, available tool support, and major concerns surrounding SBOM usage \cite{Xia2023An}. Their findings indicate limited adoption in practice. Even among projects that generate SBOMs, there is no clear consensus on what information to include, despite official recommendations and guidelines, underscoring the need for clearer expectations and stronger standardization.

These observations are consistent with evidence reported in the Linux Foundation’s SBOM readiness survey, which analyzed responses from 412 organizations worldwide \cite{LinuxFoundationSBOM2022}. The survey revealed notable gaps in organizational familiarity with SBOMs, as well as uncertainty regarding SBOM production and consumption strategies. In addition, many respondents expressed skepticism about whether SBOM requirements are being broadly adopted across the software industry, raising concerns about alignment with emerging regulatory and policy initiatives.

Beyond adoption awareness, Xia et al. further identified significant limitations in existing SBOM tooling, particularly for SBOM consumption and vulnerability handling, underscoring the need for more robust, interoperable, and enterprise-ready solutions \cite{Xia2023An}. The authors also emphasized the importance of validation and verification mechanisms to ensure the reliability and trustworthiness of SBOM data, especially given the risk of incomplete or manipulated metadata. Together, these findings highlight the need for continued improvements in SBOM tooling and practices to support meaningful adoption.

Nocera et al. focused specifically on open-source projects hosted on GitHub, uncovering a growing trend in SBOM adoption \cite{Nocera2023Software}. Despite this increase, the overall uptake remains limited, particularly in smaller projects. Their comparison of SPDX and CycloneDX formats highlighted that while SPDX is widely adopted, CycloneDX offers superior support for security-related features, which is a critical consideration for projects prioritizing security. 

One of the closest related works is the work of Yu et al., who compared the quality of SBOM generation across GitHub Dependency Graph, Syft, Trivy, and Microsoft SBOM Tool \cite{Yu2024CorrectnessSBOM}. The study covered 7,876 open-source repositories spanning nine programming languages. The authors compared the tools’ outputs by examining the number of detected packages, measuring pairwise overlap using Jaccard similarity over (name, version) dependency sets, and quantifying the presence of duplicate packages within individual SBOMs. Although this work is similar to ours, their analysis focused primarily on how different tools vary in reporting the number of dependencies in SBOMs, whereas we focused on how the tools differ in reporting key metadata, such as component version, licensing, and compliance with NTIA requirements.

Another close work to ours is Wang et al.~\cite{wang2026}.
They present an empirical study of 3,287 repositories across four languages (C, C++, Java, and Python), focusing on three dimensions: structural compliance with SBOM standards, inter-tool consistency, and information accuracy. Consistent with our results, they found inadequate compliance with policy requirements, poor inter-tool consistency, and low license reporting rates.
Our study is complementary in multiple ways.
We cover more programming languages, which enables us to draw parallels across other ecosystems. 
We also include an in-depth investigation into the potential root causes of missing license information across ecosystems and evaluate the practical utility of SBOMs for automated vulnerability tracking. 
The work of Wang et al.~\cite{wang2026} was published recently, and was conducted concurrently with our study. 
We believe that both studies offer complementary perspectives on the state of SBOM tooling.

\begin{table*}[t]
\centering
\caption{Comparison of our study with the two most closely related 
empirical SBOM studies. \xmark~denotes not addressed.}
\label{tab:related_comparison}
\resizebox{\linewidth}{!}{%
\begin{tabular}{p{4.2cm}
                >{\raggedright\arraybackslash}p{4.0cm}
                >{\raggedright\arraybackslash}p{4.0cm}
                >{\raggedright\arraybackslash}p{4.8cm}}
\toprule
\textbf{Dimension}
    & \textbf{Yu et al.~\cite{Yu2024CorrectnessSBOM}}
    & \textbf{Wang et al.~\cite{wang2026}}
    & \textbf{This Work} \\
    & \textit{DSN 2024}
    & \textit{ACM TOSEM 2026}
    & \textit{Under submission} \\
\midrule

Study scale
    & 7,876 repos; 9 languages; 4 tools
    & 3,287 repos; 3 languages; 6 tools
    & \textbf{10,000 repos; 10 languages; 4 tools} \\[6pt]

SBOM compliance assessment
    & \xmark
    & extended NTIA field set
    & all 7 NTIA minimum elements, per language \\[6pt]

Metadata completeness analysis
    & version only; pinned vs.\ unpinned
    & version, license, PURL; field-level per tool
    & version, license, PURL; top-level \& transitive, per tool \& language \\[6pt]

Inter-tool comparison
    & component counts + Jaccard similarity
    & triple-factor matching; field-level consistency
    & counts, version, license; Kruskal-Wallis + Dunn per language \\[6pt]

Ground-truth accuracy evaluation
    & Python only
    & Python only; 100 repos
    & Python, Java, JS; 200 repos each \\[6pt]

Vulnerability tracking utility
    & \xmark
    & \xmark
    & PURL-based via OSV.dev; effect of version precision on detection, per tool \& language \\[6pt]

Root-cause analysis of missing metadata
    & parser limitations \& metadata file constraints
    & standard ambiguity, scope definition
    & 7 causes at package manager level \\

\bottomrule
\end{tabular}%
}

\end{table*}

Table~\ref{tab:related_comparison} summarises the key dimensions 
along which our study differs from the two most directly related 
works, \cite{Yu2024CorrectnessSBOM} and \cite{wang2026}. The table highlights 
that while all three studies share a concern for SBOM quality and 
inter-tool comparison, each addresses a distinct set of research 
questions, and our work uniquely combines the broadest language 
coverage, a focus on the GitHub-native SBOM Tool, end-to-end 
vulnerability tracking evaluation, and a systematic root-cause 
analysis of missing metadata at the package-manager level.

\subsection{Compliance with NTIA Requirements.}
Compliance with NTIA requirements is a pivotal aspect of SBOM implementation, and several studies have explored the associated challenges and benefits. Zahan et al. identified the top five benefits and challenges of aligning SBOMs with NTIA guidelines \cite{Zahan2023Software}. Their findings suggest that while SBOMs significantly enhance transparency and security, the complexity of achieving compliance poses a substantial hurdle. Similarly, Torres et al. conducted a study through which multiple SBOM generating tools were assessed according to NTIA requirements to see how compliant they are with these latter \cite{Torres-Arias2023A}. Only 1\% of SBOMs coming from the assessed Docker images are fully NTIA compliant. 

Nocera et al. analyzed 119 SBOMs—89 using CycloneDX and 30 using SPDX—spanning 84 projects and 22 owners \cite{NOCERA2025112540}. In their assessment, the authors noted that the analyzed SPDX SBOMs contained all NTIA minimum data fields \emph{at least once}, with the supplier name and dependency relationship appearing in only a single SBOM. Based on this, they concluded that 3\% ($1/30 \times 100$) of the SPDX SBOMs satisfied all NTIA minimum data fields. 

\subsection{Challenges in SBOM Implementation.}
Research into the challenges of SBOM implementation has identified a range of technical and organizational obstacles. Stalnaker et al. detailed 12 major challenges in creating and using SBOMs, including tool deficiencies, domain-specific challenges, and difficulties in integrating SBOMs into existing workflows~\cite{Stalnaker2023BOMs}. These findings are echoed by Torres-Arias et al., who argue that the development of quality measurement mechanisms for SBOMs is crucial for enhancing their effectiveness~\cite{Torres-Arias2023A}.
The study emphasizes that without robust mechanisms to assess SBOM quality, their potential benefits in security and transparency may not be fully realized. Dalia et al. published a paper through which they conducted an analysis of the challenges holding back SBOM adoption \cite{Dalia2024}. It further conducted a comparative analysis between multiple SBOM generation tools based on the features considered, based on the challenges initially stated in the paper. Bi et al. proposed a study through which the first comprehensive classification of SBOM-relevant issues and potential solutions has been provided, along with an identification of the different phases of the SBOM lifecycle and their characteristics \cite{Bi2023On}. In this classification, issues were correlated with various stages of the SBOM lifecycle, offering a structured analysis that serves as a guide for developers. This analysis aids in understanding the challenges associated with SBOMs and offers recommendations for effectively integrating SBOMs to address development issues in real-world scenarios. Additionally, the study identifies gaps in the existing production and usage of SBOMs, highlighting shortcomings in current practices. Based on these insights, the authors suggest future research directions aimed at improving SBOM quality and adoption, providing a roadmap for advancing the field.

\subsection{Impact of SBOMs on Software Security and Transparency.}
SBOMs are increasingly recognized for their role in improving software supply chain security and transparency. Xia et al. explored the integration of blockchain technology into SBOMs as a means of enhancing security by providing a tamper-resistant mechanism for SBOM sharing \cite{Xia2024Trust}. This study suggests that blockchain can significantly improve the trustworthiness of SBOMs, particularly in environments where security is paramount. Similarly, Carmody et al. examined the potential of SBOMs in the medical technology supply chain, highlighting their ability to improve transparency and trust among stakeholders \cite{Carmody2021Building}. Sharma et al. introduced SBOM.exe, a tool designed to detect the malicious usage of dynamic features in Java by verifying the binary integrity of an application's dependencies at runtime \cite{Sharma2024}.

\section{Threats to Validity}
\label{sec:limitations}
\textbf{Generalization of the results.} Our study’s findings may not generalize to all software projects or SBOM tools. We generated 10,000 SBOMs across 10 programming languages using the GitHub SBOM tool and the three other popular tools: the Microsoft SBOM, Syft and Trivy. While this covers a variety of languages and tooling approaches, it still represents only a subset of the broader software ecosystem. Projects outside our selection—especially those in different domains or using different dependency management practices—might show different patterns. 

Moreover, we only considered open-source projects hosted on GitHub. Proprietary software or projects hosted elsewhere may have different dependency characteristics, so caution is needed when extending our conclusions to those contexts. Future studies including a wider range of languages, tools, and project types are needed to confirm whether our observations hold more generally.

\textbf{Internal threat to validity.} Potential mismatches in SBOMs could arise if different tools analyze different snapshots or branches of a repository, or if their internal generation mechanisms differ. To control for this, \mstool, Trivy, and Syft were executed on locally cloned repositories, using a snapshot of each repository’s default branch to ensure identical source code, dependency manifests, and project structures. In contrast, GitHub SBOMs were obtained via the GitHub REST API; although these are also generated from the default branch, the internal mechanisms remain opaque and may differ from those of the other tools. It is also important to note that the \mstool did not successfully generate SBOMs for 254 out of 10,000 projects (2.54\%). These failures were distributed across multiple programming languages, indicating that they were not concentrated in a single ecosystem. While this represents a small proportion of the overall dataset, the excluded repositories may slightly affect the completeness of the comparison. Nevertheless, the large remaining sample provides broad coverage across languages and helps mitigate the impact of these isolated tool limitations.

\textbf{Reproducibility limitations.}
Reproducibility in our study differs across tools due to the nature of how SBOMs are
generated. The GitHub REST API does not support generating SBOMs for historical states of
a repository, making it impossible to recreate the same GitHub-generated SBOMs at a later
time. However, our replication package includes all GitHub-generated SBOMs, which can be
used directly for reproducibility purposes. In contrast, for \mstool, \trivy,
and \syft, SBOMs can be regenerated from the same locally cloned repository snapshot,
ensuring consistent results across repeated analyses. Each generated SBOM records its
creation timestamp, further supporting reproducibility for these three tools.

\textbf{Conclusion validity.}
Throughout this study, our research questions are framed primarily around programming languages, as
we collected repositories labeled under ten different languages on GitHub. However, as
discussed in Section~\ref{sub:licensing-discussion}, real-world projects are rarely
confined to a single language: most repositories are polyglot in nature, comprising
components written in multiple languages and managed by different package
managers~\cite{tomassetti2014polyglot,wen2023multilingual}. This means that the SBOMs we analyze for a given language may contain packages drawn from
entirely different ecosystems, and the results we attribute to a language may in part reflect the behavior of those other ecosystems rather than the primary one. Practitioners and researchers should therefore interpret our per-language conclusions with this caveat in mind, as they represent the experience of repositories \textit{labeled} as a given language, not of purely single-language projects.

\section{Conclusion}
\label{sec:conclusion}
This study examined the quality of automatically generated SBOMs at scale, using GitHub’s native SBOMs as a representative case and comparing them with widely used third-party tools. Our results show that while automated SBOM generation has become easier with the availability of multiple tools, important gaps remain that limit its effectiveness for security and compliance-driven use cases.

In particular, the absence of supplier information, the incomplete and uneven version reporting introduces blind spots that propagate to transitive dependencies and directly weaken vulnerability and license analysis. 
Our findings demonstrate that missing or non-exact versions force downstream tools to either over-approximate vulnerabilities or ignore affected components entirely.

The comparison across SBOM tools further indicates that no single approach currently offers comprehensive coverage across languages and metadata dimensions. GitHub’s SBOMs provide broad dependency visibility and strong identifier support, while the \mstool prioritizes version completeness at the cost of coverage. The variability observed across ecosystems suggests that SBOM quality is shaped as much by language-specific tooling and conventions as by the SBOM generator itself.

Overall, these results suggest that automated SBOMs are best viewed as a baseline rather than a complete solution. Improving supplier attribution, version propagation, and license extraction
will be critical for SBOMs to fully support large-scale vulnerability management and supply-chain risk assessment. Our findings highlight concrete areas where SBOM tooling and standards can evolve to better align with their intended role in securing modern software supply chains.

\section*{Declarations}
\label{sec:S&D}

\noindent\textbf{Funding} We acknowledge the support of the Natural Sciences and Engineering Research Council of Canada (NSERC)[funding reference number RGPIN-2023-05163].\\
\noindent\textbf{Ethical Approval} Not applicable.\\ 
\noindent\textbf{Informed Consent}  Not applicable.\\
\noindent\textbf{Author Contributions} \\
Conceptualization: Kawsar Ahmed Bhuiyan, Diego Elias Costa.\\ 
Methodology: Kawsar Ahmed Bhuiyan, Mohamed Bilel Besbes, Rachna Raj, Diego Elias Costa.\\
Data Collection: Kawsar Ahmed Bhuiyan, Adam Al Assil.\\ 
Analysis: Kawsar Ahmed Bhuiyan, Mohamed Bilel Besbes.\\ 
Writing - original draft preparation: Kawsar Ahmed Bhuiyan, Adam Al Assil.\\
Writing - review and editing: Mohamed Bilel Besbes, Rachna Raj, Diego Elias Costa.\\
Supervision: Diego Elias Costa.\\
\noindent\textbf{Data Availability Statement} The data and code used in this study are available at \url{https://doi.org/10.5281/zenodo.18883005}.\\
\noindent\textbf{Conflict of Interest} Not applicable.\\
\noindent\textbf{Clinical Trial Number} Not applicable.

\bibliographystyle{abbrv}
\bibliography{bibliography}

\end{document}